\documentclass[global,referee,a4paper]{svjour}
\journalname{Bulletin of Mathematical Biology}
\usepackage[width=15cm,height=23cm,centering]{geometry}
\usepackage{graphicx}
\usepackage{amsmath}
\usepackage{bbm}
\usepackage{amsfonts}
\usepackage{amssymb}
\usepackage{amsbsy}
\usepackage[T1]{fontenc}
\usepackage{theorem}
\usepackage{stmaryrd}
\usepackage{tipa}
\usepackage{textcomp}
\usepackage{subfigure}
\usepackage{epsfig}
\usepackage[sort,colon]{natbib}
\usepackage{lmodern}
\usepackage{framed}
\usepackage[subfigure]{tocloft}
\usepackage{subeqnarray}
\usepackage{alphalph}
\usepackage[refpage,cfg,prefix]{nomencl}
\usepackage{multirow}
\usepackage{xifthen}
\usepackage{enumerate}
\usepackage{enumitem}
\usepackage{mathabx}
\usepackage{color}
\usepackage{wasysym}
\usepackage{bibentry}
\nobibliography*
\usepackage{verbatim}
\usepackage[hyperindex,breaklinks]{hyperref}
\hypersetup{
colorlinks=true,
urlcolor=blue,
linkcolor=black,
citecolor=black,
}
\usepackage{breakurl}
\usepackage{url}
\usepackage{setspace}
\makenomenclature

\newcommand{\ud}{\mbox{d}}

\newcommand{\bs}[1]{\boldsymbol{#1}}

\newcommand{\straightD}[2]{\frac{\ud #1}{\ud #2}}
\newcommand{\listofalgorithms}{\textbf{\Huge{List of Algorithms}}}
\newlistof{Algorithm}{exp}{\listofalgorithms}
\newcounter{instructioncounter}

\begin{document}

\title{A multi-stage representation of cell proliferation as a Markov process}
\author{Christian A. Yates\inst{1}\thanks{Corresponding author. 
E-mail: c.yates@bath.ac.uk www: www.kityates.com}
\and \qquad Matthew J. Ford\inst{3}$^,$\inst{4} \and \qquad Richard L. Mort\inst{2}}

\institute{Centre for Mathematical Biology, Department of Mathematical Sciences, University of Bath, Claverton Down, Bath, BA2 7AY, UK.
\and
Division of Biomedical and Life Sciences, Faculty of Health and Medicine, Furness Building, Lancaster University, Bailrigg, Lancaster, LA1 4YG, UK.
\and
MRC Human Genetics Unit, MRC IGMM, Western General Hospital, University of Edinburgh, Edinburgh, EH4 2XU, UK.
\and
Current Address: Rosalind and Morris Goodman Cancer Research Centre, Department of Human Genetics, McGill University, 1160 Pine Avenue West, Montreal, Quebec, H3A 1A3, Canada.}

\date{Received: date / Revised version: date}

\maketitle

\begin{abstract}
 
The stochastic simulation algorithm commonly known as Gillespie's algorithm (originally derived for modelling well-mixed systems of chemical reactions) is now used ubiquitously in the modelling of biological processes in which stochastic effects play an important role. In well-mixed scenarios at the sub-cellular level it is often reasonable to assume that times between successive reaction/interaction events are exponentially distributed and can be appropriately modelled as a Markov process and hence simulated by the Gillespie algorithm. However, Gillespie's algorithm is routinely applied to model biological systems for which it was never intended. In particular, processes in which cell proliferation is important (e.g. embryonic development, cancer formation) should not be simulated naively using the Gillespie algorithm since the history-dependent nature of the cell cycle breaks the Markov process. The variance in experimentally measured cell cycle times is far less than in an exponential cell cycle time distribution with the same mean.

Here we suggest a method of modelling the cell cycle that restores the memoryless property to the system and is therefore consistent with simulation via the Gillespie algorithm. By breaking the cell cycle into a number of independent exponentially distributed stages we can restore the Markov property at the same time as more accurately approximating the appropriate cell cycle time distributions. The consequences of our revised mathematical model are explored analytically as far as possible. We demonstrate the importance of employing the correct cell cycle time distribution by recapitulating the results from two models incorporating cellular proliferation (one spatial and one non-spatial) and demonstrating that changing the cell cycle time distribution makes quantitative and qualitative differences to the outcome of the models. Our adaptation will allow modellers and experimentalists alike to appropriately represent cellular proliferation - vital to the accurate modelling of many biological processes - whilst still being able to take advantage of the power and efficiency of the popular Gillespie algorithm.

\end{abstract}

\maketitle


\section{Introduction}

In a well-mixed solution of chemicals undergoing reactions, non-reactive collisions occur far more often than reactive
collisions allowing us to neglect the fast dynamics of motion. We can thus assume that the time between reactive collision events is exponentially distributed with rates which are a combinatorial function of the numbers of available reactants \citep{gillespie1976gmn,gillespie1977ess}. This premise is the basis of the \citet{gillespie1976gmn} stochastic simulation algorithm\footnote{Although Gillespie was amongst the first to popularise the stochastic simulation algorithm and derived it from physical considerations, its conception dates back to the work of \citet{doob1945mcd} and \citet{feller1940ide}. It was derived in a different form by \citet{kurtz1972rbs}.}. The Gillespie algorithm has become a ubiquitous algorithm for the simulation of stochastic systems in the biological sciences, in particular in computational systems biology \citep{szekely2014sss}.

\begin{figure}[h]
\begin{center} 
\includegraphics[width=0.46\textwidth]{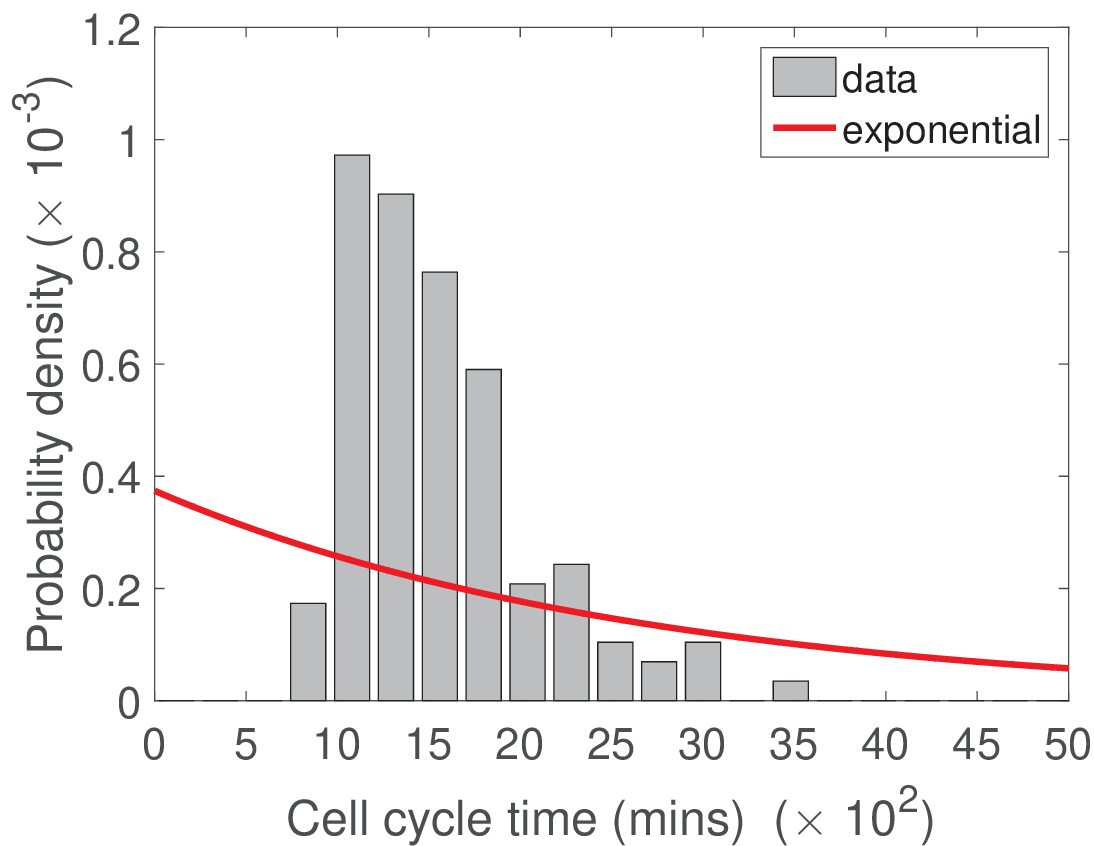}
\end{center}
\caption{The poor agreement between the best-fit exponential distribution (red curve) and experimentally determined cell cycle times in NIH 3T3 mouse embryonic fibroblasts  grown \textit{in vitro} (grey histograms). The rate parameter of the exponential distribution was fitted by minimising the sum of squared residuals between the curve and the bars of the histogram.}
\label{figure:poor_exponential_fit}
\end{figure}

However, the Gillespie algorithm is often used inappropriately to represent processes for which the inter-event time is not exponentially distributed. One prevalent example of this is in the simulation of the cell cycle \citep{baar2016smi,castellanos2014smc,ryser2016qdf,figueredo2014csd, turner2009cbc, mort2016rdm,zaider2000tcp} (see figure \ref{figure:poor_exponential_fit}).  The assumption of memorylessness, and consequently exponentially-distributed cell cycle times, means that with high probability a daughter cell may divide immediately after the division event which created it. This is not biologically plausible since each cell is required to pass through the $G_1$, $S$, $G_2$ and $M$ phases of the cell cycle before division, and these phases (in particular $S$-phase) are rate-limiting.

As an illustrative example, \citet{turner2009cbc} present a well-mixed stochastic model of small populations of cancer stem cells, which they use to suggest treatment strategies. Both analytical and simulated results pertaining to the survival of the tumour are based on the assumption that inter-division times are exponentially distributed.  We demonstrate later in this paper that using the correct cell cycle time distributions (CCTDs) could alter their results leading to different suggested treatment strategies.

Similarly, in a spatially extended context, \citet{baker2010cmf} investigate the role of spatial correlations on individual-based models of cell migration, proliferation and death, designed to represent experimental assays of cell behaviour in culture. The individual-based models they employ are on-lattice, volume-exclusion processes in two and three dimensions in which cell migration, proliferation and death events are all considered to be exponentially distributed. Cell proliferation occurs when a cell is chosen to divide, placing a daughter cell at one of its neighbouring lattice sites. \citet{baker2010cmf} demonstrate the effects of spatial correlations in these models. In particular, they find that changing the motility of the cells can alter their net rate of growth in comparison to the logistic growth predicted by a simple mean field assumption which neglects the effects of correlations. This effect is due to the fact that cell motility serves to break up correlations allowing more proliferation events to occur in comparison to the lower motility case. By explicitly considering the correlations between the occupancies of pairs of lattice sites, \citet{baker2010cmf} derive a more accurate population-level model which better represents the growth in the number of cells over time for a diverse range of parameter values. We will investigate the effects of incorporating more realistic CCTDs on the outcomes of the model simulations.

Non-Markovian simulation methods exist for events which do not have exponentially distributed inter-event times \citep{boguna2014snm}. However,  these algorithms are often difficult to understand and complex to encode since we are required to keep track of every cell individually. This presents a potential barrier to their use and consequently a barrier to the appropriate modelling of CCTDs.

Given the ubiquity of the Gillespie algorithm, it would be significantly more beneficial if we could decompose the cell cycle into a series of exponentially distributed events which could be naturally encoded in the framework of the Gillespie algorithm. One potential solution to this problem is the use of the hypoexponential family of distributions. It has been suggested that these distributions can be used to accurately represent phases of the cell cycle (and, by closedness of the sums of these distributions, the cell cycle itself) \citep{stewart2009pmc}. Hypoexponential distributions are made up of a series of $k$ independent exponential distributions, each with its own rate, $\lambda_i$, in series. If $k$ is large then these models may face issues of parameter identifiability. 

Recently, \citet{weber2014qlv} have suggested that a delayed hypoexponential distribution (consisting of three delayed exponential distributions in series) could be used to appropriately represent the cell cycle. These delayed exponential distributions represents the $G_1$, $S$ and a combined $G_2/M$ phases of the cell cycle.  Their model is an extension of the seminal stochastic cell cycle model of \citet{smith1973dcc} who use a single delayed exponential distribution to capture the variance in the cell cycle. Delayed hypoexponential distributions representing periods of the cell cycle have been justified by appealing to the work of \citet{bel2009sct}. \citet{bel2009sct} showed that the completion time for a large class of complex theoretical biochemical systems, including DNA synthesis and repair, protein translation and molecular transport, can be well approximated by either deterministic or exponential distributions. 

In this paper we consider two special cases of the general hypoexponential distribution: the Erlang and exponentially modified Erlang distribution which, in turn, are special cases of the Gamma and exponentially-modified Gamma distributions. For reference their PDFs $P_E$ and $P_{EME}$, respectively, are given below: 
\begin{equation}
P_E(x)=\frac{\lambda^k x^{k-1} e^{-\lambda x}}{(k-1)!}, \quad P_{EME}=\frac{\lambda_1^k\lambda_2e^{-\lambda_2 x}}{(k-1)!}\int_0^x e^{(\lambda_2-\lambda_1) t}t^{k-1}\ud t. \label{equation:erlang_andEM_erlang_PDFs}
\end{equation}

\begin{figure}[h]
\begin{center} 
\subfigure[]{
\includegraphics[width=0.46\textwidth]{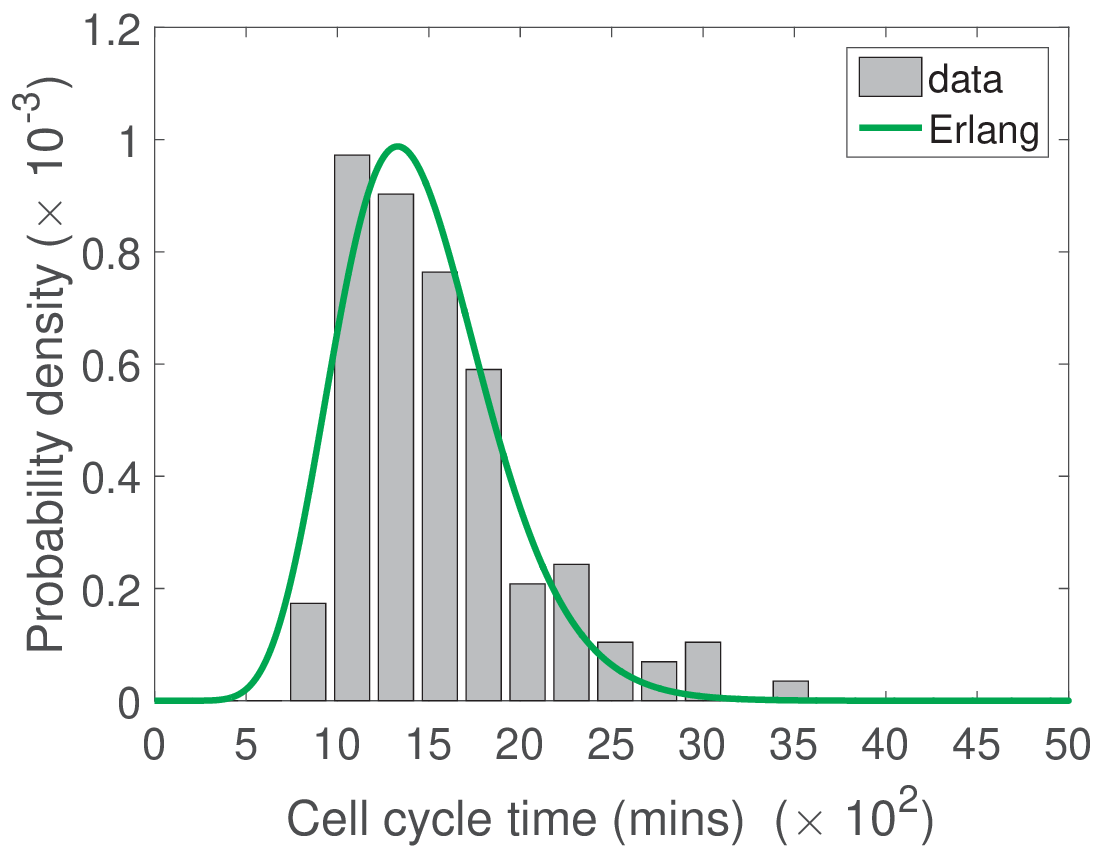}
\label{figure:data_comparison_Erlang}
}
\subfigure[]{
\includegraphics[width=0.46\textwidth]{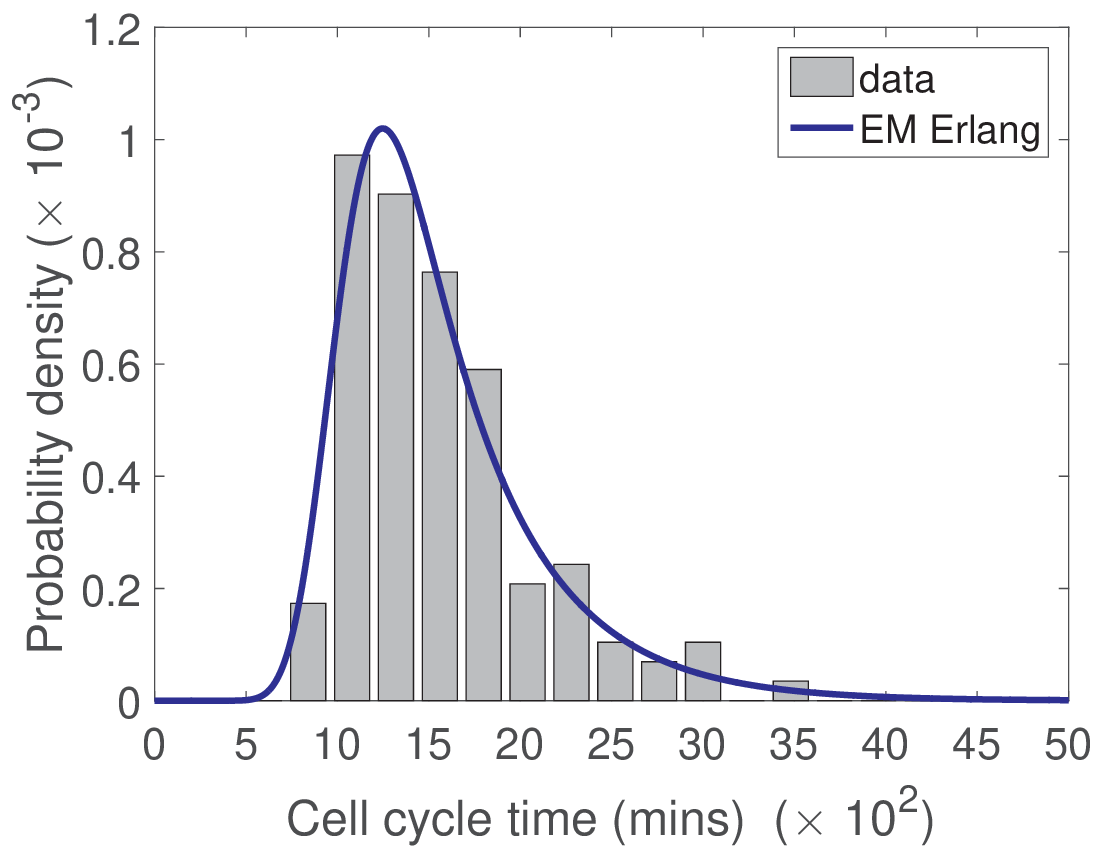}
\label{figure:data_comparison_EM_Erlang}
}
\end{center}
\caption{The agreement between experimentally determined cell cycle times in NIH 3T3 mouse embryonic fibroblasts grown \textit{in vitro} (grey histograms) and \subref{figure:data_comparison_Erlang} the Erlang distribution (green curve) \subref{figure:data_comparison_EM_Erlang} the exponentially modified Erlang distribution (blue curve).}
\label{figure:better_data_fits}
\end{figure}

With suitable parameter choices, both distributions have been shown to provide good fits to large numbers of experimentally-derived CCTDs \citep{golubev2016aie} (see Fig \ref{figure:better_data_fits} for one such example).
As special cases of the hypoexponential distributions, these distributions also have the significant advantage that they can be simulated using the ubiquitous Gillespie stochastic simulation algorithm. This will allow for the appropriate representation of CCTDs in stochastic models of cell populations, in contrast to the inappropriate exponentially distributed times which are commonly used \citep{baar2016smi,castellanos2014smc,ryser2016qdf,figueredo2014csd, turner2009cbc, mort2016rdm,zaider2000tcp}. Additionally the two and three parameters (respectively) of the Erlang and exponentially modified Erlang distributions (respectively) simplify parameter identification in comparison to more highly parametrised distributions.
These two choices (Erlang and exponentially modified Erlang distributions) are not the only non-monotone distributions which could be used to appropriately represent the cell cycle. However, they are the general, non-monotone, hypoexponential distributions with the fewest number of parameters (two for Erlang and three for exponentially modified Erlang). These features will aid parameter identifiability (few parameters) and crucially mean the distributions can be simulated using the Gillespie algorithm (hypoexponentiality), making these the most suitable distributions to consider. 

Figure \ref{figure:better_data_fits} demonstrates the improved agreement between the Erlang and exponentially modified Erlang distributions with the experimental data in comparison to the exponential distribution (\textit{c.f.} figure \ref{figure:poor_exponential_fit}). In each case the parameters of the distributions are fitted by minimising the sum of squared residuals, $\Sigma$, between the curve and the bars of the histogram\footnote{Note that even simpler methods of fitting are possible. In particular, given the first two (three) moments for the Erlang (exponentially modified Erlang) distribution, parameters can be identified by matching moments. This implies the knowledge of the whole distribution of cell cycle times is not necessary. For the Erlang distribution one would need only the mean and variance of cell cycle times. For the exponentially modified Erlang distribution the skewness would also be required.}. Clearly, for the exponential distribution (see figure \ref{figure:poor_exponential_fit}), the shape of the curve is incorrect. Consequently the exponential distribution gives a poor representation of the true CCTD with a sum of squared residuals $\Sigma=1.86\times 10^{-6}$. Evidently the Erlang distribution with parameters $\lambda=0.0083$ and $k=12$ gives a much better agreement to the experimental data (see figure \ref{figure:better_data_fits} \subref{figure:data_comparison_Erlang}), with a minimised sum of squared residuals, $\Sigma=1.23\times 10^{-7}$. Finally, the exponentially modified Erlang distribution with parameters $\lambda_1=0.0251$, $\lambda_2=0.0019$ and $k=26$ gives an even better agreement to the data\footnote{The three-parameter exponentially modified Erlang distribution may be poorly constrained for some data sets. That is to say there are several values of the parameters which give almost equally good fits to the data. Keeping the aim of carrying out stochastic simulation using the determined distribution in mind, we suggest a preference for acceptable parameter values with the smallest possible value of $k$. The larger the value of $k$ the more ``reaction channels'' the Gillespie algorithm must account for which, if rate-limiting, will significantly reduce the algorithm's performance.} with a minimised sum of squared residuals, $\Sigma=6.01\times 10^{-8}$. The exponentially modified Erlang distribution achieves a minimised sum of squared residuals which is around half that of the Erlang distribution. Never-the-less, both Erlang and exponentially modified Erlang are good candidates for fitting cell cycle time data and can both be simulated within the existing Gillespie framework, so will be considered here.

In Section \ref{section:multistage_cell_cycle_model} we begin by outlining a general hypoexponential model of the cell cycle and noting that many previous models of the cell cycle are special cases. By simplifying the model further we demonstrate that the Erlang and exponentially modified Erlang distributions are also special cases. 
In Section \ref{section:identical_rates}, we consider the special case of the Erlang distributed CCTD in more detail. Undertaking some simple analysis we derive the expected behaviour of the mean cell number in the case of Erlang CCTDs and demonstrate analytically that significant differences can arise in comparison to models in which exponentially distributed CCTDs are used. In Section \ref{section:illustrative_examples} we demonstrate the utility of our new CCTD representation in stochastic simulations of two biological models in which cellular proliferation is of critical importance. In each case we show, through simulation, that there are important quantitative and qualitative differences between models which represent cell cycle times appropriately and those which do not. We conclude in Section \ref{section:discussion} with a short discussion on the implications of our findings.


\section{Multi-stage model of the cell cycle}\label{section:multistage_cell_cycle_model}

We divide the cell cycle (with mean length $C$) into $k$ stages\footnote{Note, these stages do not necessarily correspond to the traditional ($G_1$, $S$, $G_2$ and $M$) phases of the cell cycle. Indeed, there is good evidence that at least one of the phases is not exponentially distributed \citep{hahn2009qac}. Rather they are arbitrary division of the cell cycle which will allow us to recreate the correct CCTD.}. The time to progress through each of these stages is exponentially distributed with mean $\mu_i$. We can represent the progression through these stages of the cell cycle as the following chain of `reactions'
\begin{equation}
 X_1\stackrel{\lambda_1}{\rightarrow} X_2\stackrel{\lambda_2}{\rightarrow} \dots \stackrel{\lambda_{k-1}}{\rightarrow} X_k \stackrel{\lambda_k}{\rightarrow} 2 X_1,\label{equation:exponential_reaction_chain}
\end{equation}
where $\lambda_i=1/\mu_i$.

The CCT under this model is hypoexponentially distributed. Although there is no simple closed form for the probability density function of the hypoexponential distribution we can find simple expressions for its mean and variance. The mean is given by the sum of the means of the exponentially distributed stage times $\sum^k_{i=1}\mu_i=C$ and the variance is the sum of the variances of these stage times, $\sum^k_{i=1}\mu^2_i$. By increasing the number of exponentially distributed stages, whilst decreasing their mean duration (in order to maintain the correct mean CCT), we can arbitrarily decrease the variance of the CCT. Many multi-stage models of the cell cycle are special cases of this general model \citep{golubev2016aie,hawkins2007mir,hillen2010fcp,hoel1974egt,kendall1948rvg,leander2014dec,leon2004gmf,nakaoka2014dmt,powell1955sfg,smith1973dcc,weber2014qlv,zilman2010sml}.

We can analyse the cell cycle reaction chain \eqref{equation:exponential_reaction_chain} further by considering the associated probability master equation (PME). Let $P(x_1,x_2,\dots,x_k,t)$ be shorthand for the probability that there are $x_1$ cells in stage one, $x_2$ in stage two and so on. The PME is
\begin{align}
 \straightD{P(x_1,x_2,\dots,x_k,t)}{t}&=\displaystyle\sum^{k-1}_{i=1}P(x_1,\dots,x_i+1,x_{i+1}-1,\dots,x_k,t)(x_i+1)\lambda_i\nonumber\\
&+P(x_1-2,x_2,\dots,x_k+1,t)(x_k+1)\lambda_k\nonumber\\
&-\displaystyle\sum^{k}_{i=1}P(x_1,\dots,x_i,x_{i+1},\dots,x_k,t)x_i\lambda_i.\label{equation:probability_master_equation}
\end{align}

By multiplying the PME by $x_j$ and summing over the state space we can find the evolution of the mean number of cells, $M_j=\sum_{\bs{x}}x_j P$, in each stage, where $\sum_{\bs{x}}$ is shorthand for $\sum^{\infty}_{x_1=1}\cdots\sum^{\infty}_{x_k=1}$ and $P$ is shorthand for $P(x_1,x_2,\dots,x_k,t)$. Upon simplification we find the following evolution equations for the mean number of cells in each stage
\begin{equation}
    \straightD{M_j}{t}=\begin{cases} 2\lambda_{k}M_{k}-\lambda_1 M_1, & \text{for}\quad  j=1, 
\\ \lambda_{j-1}M_{j-1}-\lambda_{j}M_{j}, & \text{for}\quad  j\neq 1. \end{cases} \label{equation:mean_evolution}
 \end{equation}

\section{Identical rates of progression}\label{section:identical_rates}
 
 The hypoexponential model's generality is also a significant drawback since it hampers parameter identifiability \citep{weber2014qlv}. As such, we seek to reduce the number of free parameters in the model whilst maintaining its ability to accurately represent CCTDs. Several authors have suggested using the Gamma distribution to model CCTDs \citep{hawkins2007mir,kendall1948rvg,nakaoka2014dmt,zilman2010sml}. If we assume that all transition rates, $\lambda_i$, are identically equal to $\lambda_1$ (for $i=1,\dots, k$) in our general hypoexponential model, then the time to progress through the whole cell cycle is distributed according to the sum of $k$ identically exponentially distributed random variables. It is straightforward to show (using moment generating functions or convolutions) that the CCTD, is Erlang distributed with scale parameter $\mu=C/k$ and shape parameter $k$. In analogy with the general hypoexponential case, if we decrease $\mu$ and simultaneously increase $k$ so that $\mu k = C$ remains constant, the Erlang distribution approaches the Dirac delta function centred on $C$, demonstrating that we can still arbitrarily reduce the variance to match the distribution we are trying to model.
 
 \subsection{Analysis of the CCDT with equal rates of progression}
 
We now analyse this CCTD model with identical rates of progression, noting that \citet{kendall1948rvg} studied this case extensively and we draw on some of his analyses below. 
Although for this special case it is possible to derive a closed form first order partial differential equation for the evolution of the generating function corresponding the the master equation \eqref{equation:probability_master_equation}, solving the associated characteristic equations is analytically intractable for all but the simplest case ($k=1$) \citep{kendall1948rvg}. Instead we will focus on the mean behaviour of the cell population with which we can make some analytical progress.

In the equal rates case we can sum the individual equations in system \eqref{equation:mean_evolution} to give
 \begin{equation}
  \straightD{M}{t}=\lambda_1 M_k\label{equation:means_summed},
 \end{equation}
where $M=\sum_i^k M_i$. 

Consider the naive one stage (\textit{i.e.} $k=1$) cell-cycle mode with mean cell cycle time $C$:
\begin{equation}
 X\stackrel{1/C}{\rightarrow} 2X.\label{equation:one_step_exponential_doubling}
\end{equation}
The evolution of the mean number of cells is given by the special case of equation \eqref{equation:means_summed}: 
\begin{equation}
   \straightD{M}{t}=M/C.\label{equation:mean_k=1}
\end{equation}
In the multi-stage model, under the assumption that all cells are evenly distributed between the stages (\textit{i.e.} $M_i=M/k$), we can replace $M_k$ with $M/k$ in equation \eqref{equation:means_summed} to give a closed equation for the evolution of the total number of cells which matches equation \eqref{equation:mean_k=1}:
\begin{equation}
   \straightD{M}{t}=\lambda_1 M/k=M/C.\label{equation:means_summed_even_assumption}
\end{equation}

\begin{figure}[h]
\begin{center} 
\subfigure[]{
\includegraphics[width=0.46\textwidth]{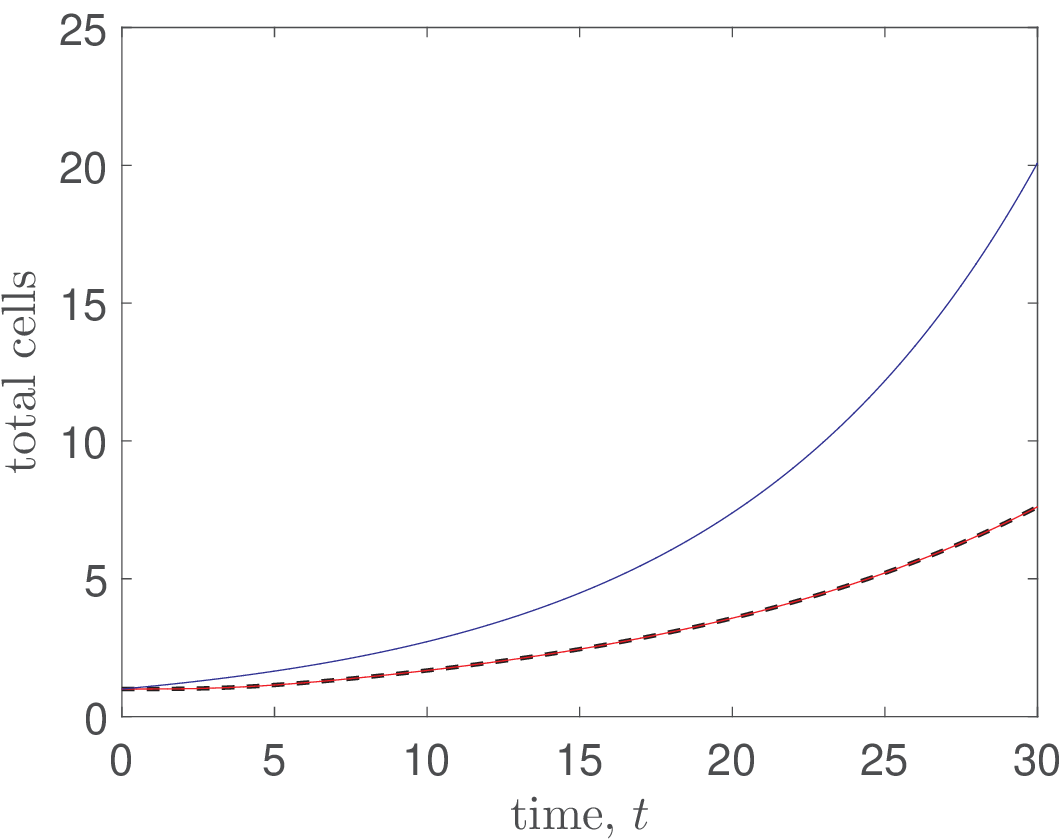}
\label{figure:stages_summed}
}
\subfigure[]{
\includegraphics[width=0.46\textwidth]{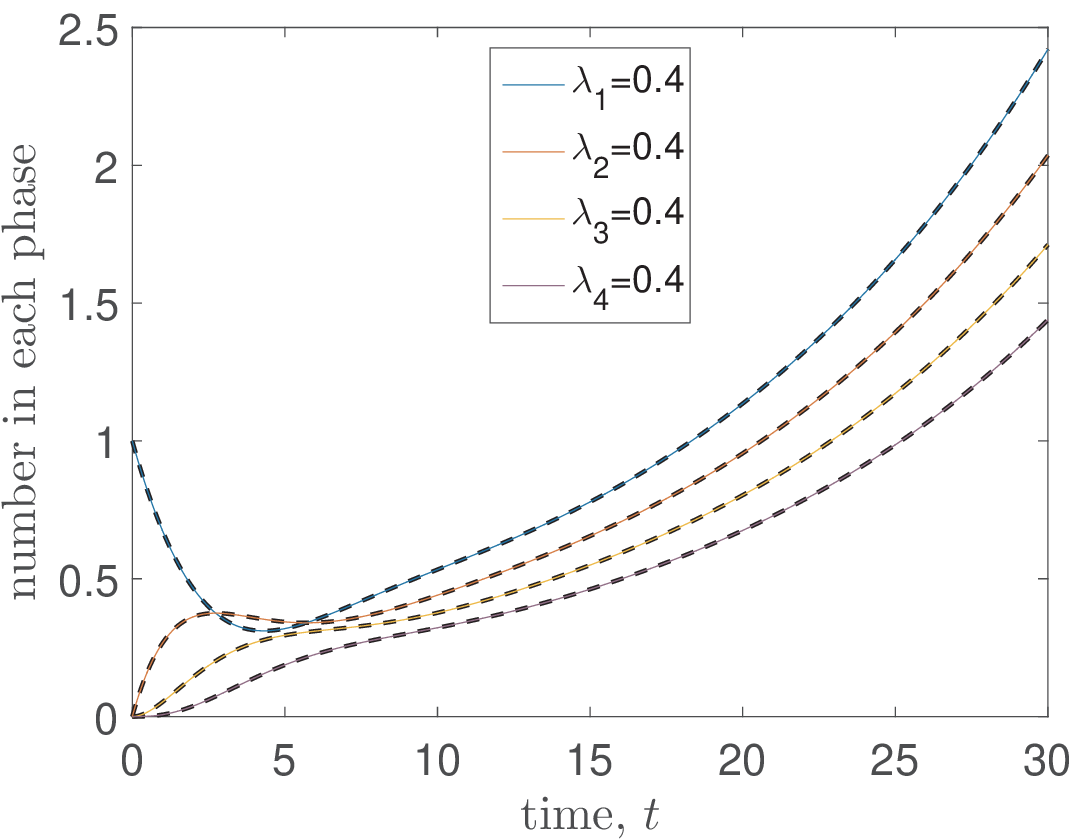}
\label{figure:all_four_stages}
}
\end{center}
\caption{The evolution of the of \subref{figure:stages_summed} the total numbers of cells  and \subref{figure:all_four_stages} the mean numbers of cells in each of $k=4$ stages.
In panel \subref{figure:stages_summed} we plot the numerical (red line) and analytical (dashed black line) solutions for the total mean number of cells in the case $k=4$ and according to the naive ($k=1$) cell cycle model (analytical solution of equation \eqref{equation:mean_k=1} - blue line). In panel \subref{figure:all_four_stages} analytically determined solutions (see equations \eqref{equation:analytical_mean_k=4_m=1}--\eqref{equation:analytical_mean_k=4_m=4}) are plotted as dashed black lines and their numerical counterparts on top as solid coloured lines. The average CCT is $C=10$ arbitrary time units. The average period of each stage is equal ($\mu_i=C/k=2.5$, $\lambda_i=1/\mu_i=0.4$ for $i=1,\dots,k$).}
\label{figure:analytical_solutions}
\end{figure}

However, the assumption on the even distributions of cells between stages is incorrect. This leads to differences not just, as might be expected, between the variation exhibited by the multi-stage and single-stage models, but also between their mean behaviour. In figure \ref{figure:analytical_solutions} \subref{figure:stages_summed} a clear difference between the $k=1$ and $k=4$ models is evident. The mean total cell number grows significantly more slowly in the $k=4$ case than the $k=1$ case. This is true for all models in which $k>1$. Intuitively, exponentially distributed CCTs imply that the most probable time for a cell to divide is the current time. Once a cell has divided it is immediately able to divide again with high probability allowing cells proliferating under the exponentially distributed CCT assumption to reinforce their numbers. This is in direct contrast to cells with Erlang distributed CCTs (with the same mean but $k>1$) which, with high probability, will wait for a period of time before dividing. In short, the larger variance of the exponentially distributed CCT population allows it to grow more rapidly.

Introduced the scaled variables $m_j=M_je^{kt/C}$, for $j=1,\dots, k$.
Under the assumption of  identical transition rates, equation system \eqref{equation:mean_evolution} can be reduced to a closed equation for the scaled mean number of cells in the $k^{th}$ stage
\begin{equation}
 \frac{\ud^k m_k}{\ud t^k}=2\left(\frac{k}{C}\right)^k m_k,
\end{equation}
and a set of $k-1$ ODEs which relate the number of cells in the other stages to $m_k$
\begin{equation}
 m_j=\left(\frac{C}{k}\right)^{k-j}\frac{\ud^{k-j} m_k}{\ud t^{k-j}}, \quad \text{for} \quad j=1,\dots, k-1.
\end{equation}
Under the given initial conditions, a single cell in the first stage and no cells in any other stages, we can solve these equations to find the unscaled mean number of cells in each stage:
\begin{equation}
 M_j=\frac{2^{(1-j)/k}}{k}\sum_{r=0}^{k-1}z^{(1-j)r}\exp((2^{1/k}z^r-1)k t/C)\label{equation:analytical_cells_in_each_stage},
\end{equation}
where $z$ is the first $k^{th}$ root of unity \citep{kendall1948rvg}. Although this expression looks complicated in some cases it is possible to express $M_j$ in a simple closed form. For example, when $k=4$
\begin{align}
 M_1=&\frac{\exp(-4t/C)}{2}\left\{\cosh\left(\frac{2^{9/4}t}{C}\right)+\cos\left(\frac{2^{9/4}t}{C}\right)\right\}\label{equation:analytical_mean_k=4_m=1},\\
 M_2=&\frac{\exp(-4t/C)}{2^{5/4}}\left\{\sinh\left(\frac{2^{9/4}t}{C}\right)+\sin\left(\frac{2^{9/4}t}{C}\right)\right\}\label{equation:analytical_mean_k=4_m=2},\\
  M_3=&\frac{\exp(-4t/C)}{2^{3/2}}\left\{\cosh\left(\frac{2^{9/4}t}{C}\right)-\cos\left(\frac{2^{9/4}t}{C}\right)\right\}\label{equation:analytical_mean_k=4_m=3},\\
 M_4=&\frac{\exp(-4t/C)}{2^{7/4}}\left\{\sinh\left(\frac{2^{9/4}t}{C}\right)-\sin\left(\frac{2^{9/4}t}{C}\right)\right\}\label{equation:analytical_mean_k=4_m=4}.
\end{align}
A comparison between these analytical solutions and their numerically solved counterparts demonstrates their veracity in Figure \ref{figure:analytical_solutions} \subref{figure:all_four_stages}.

By summing equation \eqref{equation:analytical_cells_in_each_stage} over all values of $j=1,\dots,k$ we can also find an expression for the total number of cells in a population:
\begin{equation}
 M(t)=\frac{1}{2k}\sum_{r=0}^{k-1}\frac{2^{1/k}z^r}{2^{1/k}z^r-1}\exp\left((2^{1/k} z^r -1)kt/C\right)\label{equation:analytical_total_population}.
\end{equation}

Although these formulae (equations \eqref{equation:analytical_mean_k=4_m=1}--\eqref{equation:analytical_total_population}) may be useful in specific cases where the closed form of the solution is readily accessible, their real utility is in shedding light on the asymptotic properties of the mean numbers of cells. 

In the limit that $t$ gets large for finite $k$ the dominant term in the summation in equation \eqref{equation:analytical_cells_in_each_stage} corresponds to the case $r=0$. Indeed for $k\leq 28$ the real parts of the other elements in the summation are negative and hence these terms decay \citep{kendall1948rvg}. Thus we have
\begin{equation}
 \lim_{t\rightarrow\infty}M_j\approx\frac{1}{k}2^{(1-j)/k}\exp\left(t \alpha_k/C\right)\label{equation:t_infinity_limit_stages},
\end{equation}
where
\begin{equation}
 \alpha_k= k\left(2^{1/k}-1\right).
\end{equation}
Summing equation \eqref{equation:t_infinity_limit_stages} over all $k$ stages leads to the asymptotic behaviour of the cell population as a whole:
\begin{equation}
 \lim_{t\rightarrow\infty}M(t)=\frac{2^{1/k}}{2\alpha_k}\exp\left(t \alpha_k/C\right)\label{equation:t_infinity_limit_total}.
\end{equation}
Equation \eqref{equation:t_infinity_limit_total} can be re-written as
\begin{equation}
 \lim_{t\rightarrow\infty}M(t)=\frac{2^{1/k}}{2\alpha_k}\left( e^{\alpha_k}\right)^{\left(t/C\right)}\label{equation:t_infinity_limit_total_rewritten}.
\end{equation}
For all $k>1$, not only is the base of the exponent $t/C$ less then $e$ (since $\alpha_k<1$, for $k>1$), but the coefficient is less than unity \citep{kendall1948rvg}. This implies that, asymptotically, the expected total number of cells in a multi-stage model will always be less than the number expected in a single stage cell cycle model (which can be determined upon substituting $k=1$ in to \eqref{equation:t_infinity_limit_total}).

Note that in the limit as $k\rightarrow \infty$, $\alpha_k\rightarrow\ln 2$. 
Thus, as might have been expected for the deterministic model resulting from the limit $k\rightarrow \infty$, the asymptotic population grows with base $2$, doubling at regular intervals as the cells divide synchronously:
\begin{equation}
 \lim_{ k\rightarrow \infty}\lim_{t\rightarrow \infty} M=\lim_{ k\rightarrow \infty} \frac{2^{1/k}}{2\alpha_k}\cdot 2^{kt/C }.\label{equation:part_limit}
\end{equation}
Surprisingly though, the coefficient of $2^{k/C t}$ does not tend to unity in equation \eqref{equation:part_limit} as might have been expected. Thus the total population grows like
\begin{equation}
 \lim_{k\rightarrow \infty}\lim_{t\rightarrow \infty} M\approx 0.721\cdot 2^{kt/C }.\label{equation}
\end{equation}

Reversing the order of limits and taking the limit as $k$ tends to infinity of equation \eqref{equation:analytical_total_population} for finite $t$ gives the limit
\begin{equation}
 \lim_{k\rightarrow\infty}M(t)=2^{\lfloor t/C \rfloor},\label{equation:k_infinity_limit_total_non_integer}
\end{equation}
for non-integer value of $t/C$, where $\lfloor x\rfloor$ gives the integer part of $x$ \citep{kendall1948rvg}. For integer values of $t/C$ the limit is 
\begin{equation}
  \lim_{k\rightarrow\infty}M(t)=\frac{3}{4} 2^{\lfloor t/C \rfloor}\label{equation:k_infinity_limit_total_integer},
\end{equation}
corresponding to the algebraic mean of the limits of equation \eqref{equation:k_infinity_limit_total_non_integer} as integers values are approached from the left and right hand sides. This ``deterministic'' doubling process is unsurprising since the waiting time distribution tends to a delta function in the $k\rightarrow \infty$ limit, implying that cell division is synchronous.

\subsection{Cells are not distributed proportional to stage length}

Returning to equations \eqref{equation:mean_evolution} under the assumption of identical rates of progression through the stages, we can derive corresponding equations for the mean proportion of cells in each stage, $\hat{M}_j=M_j/M$ for $j=1,\dots, k$:
\begin{equation}
    \straightD{\hat{M}_j}{t}=\begin{cases} \lambda_1 \left(2\hat{M}_{k}-\hat{M}_1-\hat{M}_1\hat{M_k}\right), & \text{for}\quad  j=1, \\ 
    \lambda_1 \left(\hat{M}_{j-1}-\hat{M}_{j}-\hat{M}_j\hat{M}_k\right), & \text{for}\quad  j\neq 1. \end{cases} \label{equation:mean_proportion_evolution}
 \end{equation}
At steady state we have the following recurrence relations for the mean proportion of cells in each stage
 \begin{equation}
    \hat{M}^{st}_j=\begin{cases} \frac{2\hat{M}^{st}_{k}}{1+\hat{M}^{st}_{k}}, & \text{for}\quad  j=1, \\ 
    \frac{\hat{M}^{st}_{j-1}}{1+\hat{M}^{st}_k}, & \text{for}\quad  j\neq 1. \end{cases} \label{equation:mean_steady_state_proportions}
 \end{equation}
In particular, this implies that $\hat{M}^{st}_j<\hat{M}^{st}_{j-1}$ for $j=2\dots k$, so that, at steady state, as we progress through the stages we will have successively fewer cells in each stage on average (independent of the initial distribution of cells amongst different stages). 
By solving these recurrence formulae we can find exact expressions for the steady state proportions:
\begin{equation}
\hat{M}^{st}_j = (\sqrt[k]{2})^{k-j}(\sqrt[k]{2}-1) \label{equation:steady_state_proportions}.
\end{equation}
In particular, note that
\begin{equation}
\frac{\hat{M}^{st}_1}{\hat{M}^{st}_k}\rightarrow 2 \quad \text {as} \quad k\rightarrow \infty.
\end{equation}
That is to say there are twice as many cells in the first stage as the last stage at steady state when the number of stages is large.

These differences are potentially important for determining average CCTs experimentally. One popular method for determining cell cycle times is to label $S$-phase cells using two sequentially administered distinct DNA specific labels \citep{wimber1963tdl,bokhari1992ndt}. The administration of the labels is separated by a known time period. By counting cells labelled with one or both labels, and with reference to the known time period of separation, it is possible to calculate the mean duration of the $S$-phase. Once the proportion of cells in $S$-phase and the mean duration of $S$-phase have been determined it is also possible to calculate the mean cell cycle time for the population \citep{nowakowski1989bid}. 

The method outlined above implicitly makes the assumption that the number of cells in a particular phase of the cell cycle is proportional to the length of that phase. For the multi-stage model we have demonstrated that in the large time limit this is unequivocally not the case. Equation \eqref{equation:analytical_cells_in_each_stage} can also be used to show that this phenomenon holds dynamically, although the mathematics is cumbersome. Instead we solve equations \eqref{equation:mean_evolution} numerically. Numerical solution also allows us to investigate the more general hypoexponential CCTD model \eqref{equation:exponential_reaction_chain} for which no analytical solutions are available. Figures \ref{figure:mean_evolution} \subref{figure:even_unnormalised} and \subref{figure:even_normalised} display the evolution of the mean numbers and proportions (respectively) of cells in each stage for equal stage progression rates, $\lambda_1$, and figures \ref{figure:mean_evolution} \subref{figure:random_unnormalised} and \subref{figure:random_normalised} display the equivalent for unequal progression rates. The number/proportion of cells in each stage is not proportional to the mean duration of the stage, $\mu_i$, either at steady state (compare actual steady state proportions with the stars representing the normalised mean stage durations, $\mu_i/\sum \mu_i$, in \ref{figure:mean_evolution} \subref{figure:even_normalised} and \ref{figure:mean_evolution} \subref{figure:random_normalised}) or dynamically.

\begin{figure}[h!!!!!!!!!!]
\begin{center} 
\subfigure[]{
\includegraphics[width=0.46\textwidth]{./figures/mean_evolution/determinstic_trajectories_even_rates_k=5_NOT_normalised.eps}
\label{figure:even_unnormalised}
}
\subfigure[]{
\includegraphics[width=0.46\textwidth]{./figures/mean_evolution/determinstic_trajectories_even_rates_k=5_normalised.eps}
\label{figure:even_normalised}
}
\subfigure[]{
\includegraphics[width=0.46\textwidth]{./figures/mean_evolution/determinstic_trajectories_random_rates_k=5_NOT_normalised.eps}
\label{figure:random_unnormalised}
}
\subfigure[]{
\includegraphics[width=0.46\textwidth]{./figures/mean_evolution/determinstic_trajectories_random_rates_k=5_normalised.eps}
\label{figure:random_normalised}
}
\end{center}
\caption{The evolution of the of the mean numbers (\subref{figure:even_unnormalised},\subref{figure:random_unnormalised}) and proportions (\subref{figure:even_normalised},\subref{figure:random_normalised}) of cells in each of $k=5$ stages.  In all cases the average CCT is $C=10$ arbitrary time units. Panels \subref{figure:even_unnormalised} and \subref{figure:even_normalised} represent the scenario in which rates are chosen so that the average period of each stage is equal ($\mu_i=C/k=2$, $\lambda_i=1/\mu_i=0.5$ for $i=1,\dots,k$), \subref{figure:random_unnormalised} and \subref{figure:random_normalised} the scenario in which the mean stage durations, $\mu_i$ are chosen by partitioning the total CCT uniformly at random into $k=5$ parts. Rates are given by $\lambda_i=1/\mu_i$ for $i=1,\dots,k$. Stars at $t=30$ on \subref{figure:even_normalised} and \subref{figure:random_normalised} indicate the expected proportion of cells at steady state if numbers of cells in each stage were proportional to cell stage duration. For the equal progression rates model all stars overlap in \subref{figure:even_normalised}.}
\label{figure:mean_evolution}
\end{figure}

\subsection{The exponentially modified Erlang distribution}

Although the identical-stage model, which gives rise to the Erlang distribution for CCTDs, is convenient from a mathematical perspective, it has been shown to have been outperformed by a number of other distributions \citep{golubev2010emg,golubev2016aie}. In particular, by considering 77 independent CCT data sets, \citet{golubev2016aie} has recently shown that one of the most appropriate distributions for representing CCTDs is the exponentially modified Gamma (EMG) distribution. For our purposes we will require that the shape parameter of the Gamma distribution is be integer valued so that the CCTD is actually an exponentially modified Erlang (EME). This will mean that we can continue to simulate CCTs using a series of exponentially distributed random variables (albeit one of them will have a different rate). Consequently this will allow us to continue to appropriately simulate processes in which cell division is important using the popular Gillespie algorithm.
In order to generate EME distributed CCTs we modify our multi-stage cell cycle model as follows:

\begin{equation}
 X_1\stackrel{\lambda_1}{\rightarrow} X_2\stackrel{\lambda_1}{\rightarrow} \dots \stackrel{\lambda_1}{\rightarrow} X_k \stackrel{\lambda_1}{\rightarrow} X_{k+1} \stackrel{\lambda_2}{\rightarrow} 2 X_1.\label{equation:EME_reaction_chain}
\end{equation}
Note that, in system \eqref{equation:EME_reaction_chain}, the rate of progression is identical through each of the initial $k$ stages of cell cycle and that we have added an additional exponentially distributed stage at the end whose rate, $\lambda_2$, is distinct from the rate, $\lambda_1$, of the previous $k$ stages.

We can ascertain the probability density function for the EME distribution, $P_{EME}(t)$, by convolving the Erlang ($P_{ER}(t)$) and exponential ($P_E(t)$) distributions as follows, 
\begin{equation}
 P_{EME}(t)=P_E(t)\ast P_{ER}(t)=\int^t_0 \lambda_2\exp(-(t-u)\lambda_2)\cdot\frac{u^{k-1}\exp(-\lambda_1 u)\lambda_1^k}{(k-1)!}\ud u,\label{equation:EME_convolution}
\end{equation}
where $\lambda_2$ is the rate of the exponential distribution with which we are convolving and, as before, $\lambda_1$ is the rate of progression through each of the $k$ identical exponentially distributed stages which comprise the Erlang distribution.
We can simplify expression \eqref{equation:EME_convolution} to the following formulation
\begin{equation}
 P_{EME}(t)=\lambda_2 e^{-t\lambda_2}\left(\frac{\lambda_1}{\lambda_1-\lambda_2}\right)^k\left\{1-\frac{\Gamma(k,L t)}{(k-1)!}\right\},\label{equation:EME_PDF}
\end{equation}
where $L=\lambda_1-\lambda_2$ and $\Gamma(k,L t)=\int^{\infty}_{L t}z^{k-1}e^{-z}\ud z$ is the complementary incomplete gamma function.

We note that it is almost as simple to simulate this more general distribution in the Gillespie algorithm using a series of exponentially distributed stages as it is to simulate the distribution with constant rates of progression between stages. Indeed the simulation of any hypoexponential CCTD is straightforward in the Gillespie algorithm. However, the addition of extra parameters hampers their identifiability when fitting to experimental data and as such we only suggest using the Erlang or exponentially modified Erlang distributions in models of the CCTD. 

In the following section we illustrate the importance of incorporating non-exponentially distributed CCTDs into stochastic simulations of cellular proliferation. For ease of understanding we concentrate purely on Erlang CCTDs, and note that the parameters are not based on fitted CCTDs but merely chosen for illustrative purposes.

\section{Illustrative examples}\label{section:illustrative_examples}
In this section we recapitulate results from two different models which each assume exponentially distributed CCTs. The first is a well-mixed model of cancer stem cell proliferation and differentiation in a brain tumour. The second a spatially extended model of cell migration and proliferation mimicking a growth-to-confluence experimental assay. 
In each case we alter the CCTD in order to see what effect this has on the qualitative and quantitative results presented in the papers. For clarity we will restrict ourselves to Erlang distributed CCTs, but note that results are qualitatively similar for exponentially modified Erlang distributed CCTs.

\subsection{Cancer stem cell maintenance}
\citet{turner2009cbc} investigate the role of sub-populations of cells within a brain tumour possessing stem cell-like properties and responsible for maintaining the tumour. In situations (e.g. post treatment) in which there are small numbers of stem cells they consider a stochastic model of cell proliferation and differentiation. Stem cells, $S$, can undergo symmetric division in which the daughter cells possess the same characteristics as the parent cells (see equation \eqref{equation:symmetric_division}) and the stem cell population increases. They can also undergo asymmetric self renewal in which one stem cell and one progenitor cell, $P$, are produced (see equation \eqref{equation:assymetric_self_renewal}) or symmetric proliferation in which two progenitor cells result from a stem cell division (see equation \eqref{equation:symmetric_proliferation}). 
Cell cycle times are exponentially distributed with rate $\rho_s$ and fate choices (about which division type to undergo) are made at the point of division. With probability $r_1$ symmetric division occurs and with probability $r_3$ symmetric proliferation occurs. Consequently with probability $r_2=1-r_1-r_3$ asymmetric self renewal occurs. These divisions with their effective rates are captured in the reaction system \eqref{equation:symmetric_division}--\eqref{equation:symmetric_proliferation}:
\begin{align}
 S&\stackrel{\rho_s r_1}{\rightarrow}S+S,\label{equation:symmetric_division}\\
S&\stackrel{\rho_s r_2}{\rightarrow} S+P,\label{equation:assymetric_self_renewal}\\
S&\stackrel{\rho_s r_3}{\rightarrow} P+P.\label{equation:symmetric_proliferation}
\end{align}

Under the assumption of exponential CCTDs, \citet{turner2009cbc} write down and solve a simple probability master equation for the stem cell population. In particular, they consider the case in which $r_1>r_3$ which implies a positive net growth rate of the stem cell population. Under this assumption the mean number of cells in the stem cell population and variance can be shown to increase exponentially. Since the model is linear, by appealing to the central limit theorem, \citet{turner2009cbc} argue that, for large enough cell populations, the exact mean field equations given by 
\begin{equation}
\straightD{S}{t}=\rho_s(r_1-r_3)S \label{equation:turner_mean_field}
\end{equation}
will approximate the stochastic dynamics well.

\begin{figure}[h]
\begin{center} 
\subfigure[]{
\includegraphics[width=0.30\textwidth]{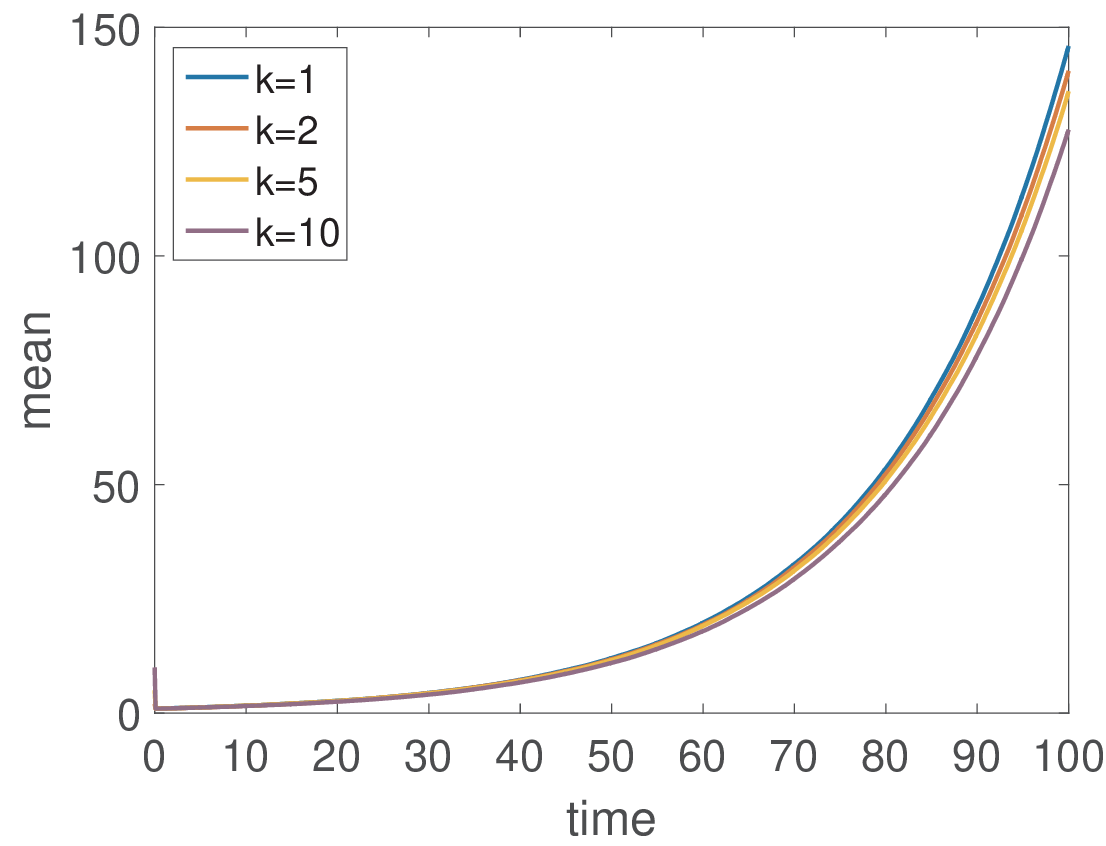}
\label{figure:mean_cell_numbers}
}
\subfigure[]{
\includegraphics[width=0.30\textwidth]{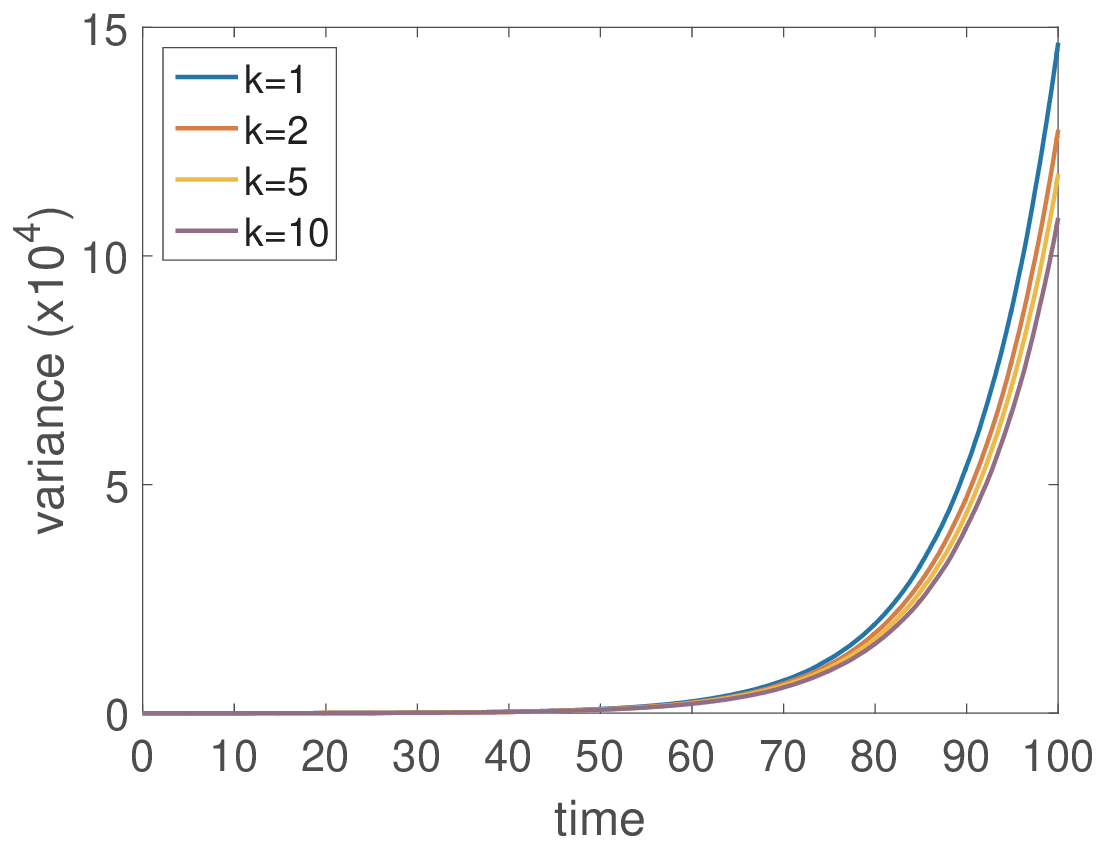}
\label{figure:variance_of_cell_numbers}
}
\subfigure[]{
\includegraphics[width=0.30\textwidth]{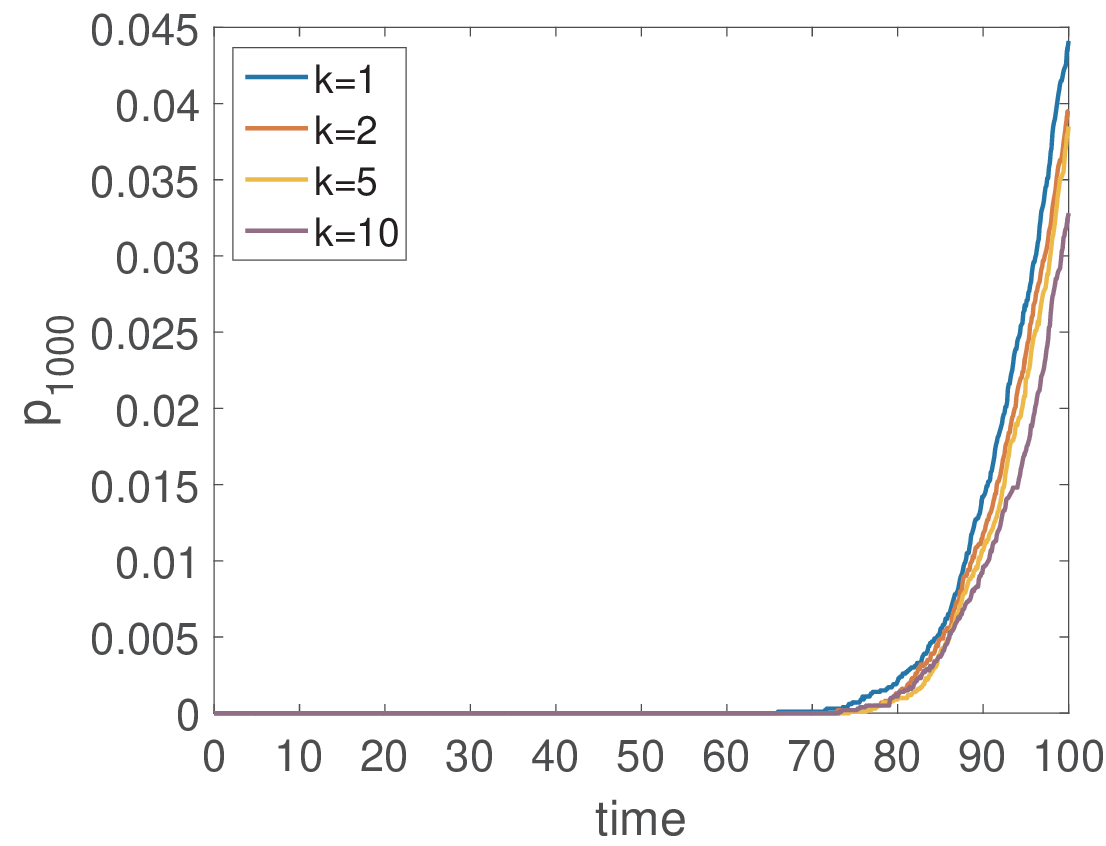}
\label{figure:p_exceed}
}
\end{center}
\caption{The evolution of the of the mean numbers, \subref{figure:mean_cell_numbers}, variance, \subref{figure:variance_of_cell_numbers}, and probability of the tumour having more than 1000 stem cells, \subref{figure:p_exceed}, for cancer stem cells under model \eqref{equation:symmetric_division}--\eqref{equation:symmetric_proliferation} with varying numbers of exponentially distributed stages of the cell cycle, $k=1,2,5,10$. The average CCT is $\rho_s=1$ and division probabilities are $r_1=0.2$ and $r_3=0.15$. Results are calculated from $M=10,000$ repeat simulations.}
\label{figure:turner_mean_variance_survival}
\end{figure}

In order to ensure a more realistic representation of the CCTD we introduce a multi-stage cell division process, as suggested above, so that the CCTDs are now Erlang distributed with the same mean $\rho_s$, but with shape parameter $k$ (and thus scale parameter $\mu_1=1/(\rho_s k)$). As, before, at each division event, a choice about the type of division to occur is made with the same probabilities ($r_1,r_2, r_3$) as previously specified. Although we still get exponential increases in the mean and variance, the rate of increase is significantly decreased (see Figs. \ref{figure:turner_mean_variance_survival} \subref{figure:mean_cell_numbers} and \subref{figure:variance_of_cell_numbers}). Crucially this means that the deterministic mean-field model derived from the original process will significantly overestimate the number of cancer stem cells in the tumour which could have significant therapeutic implications.

Under this model with $r_1>r_3$, if a tumour is not completely eradicated by treatment it is possible that it can return. It may, therefore be informative to know the probability that a tumour will reach a certain size  by a particular time in order to plan appropriate follow-up therapeutic intervention. For example we may be interested to know the evolution of the probability that the tumour has reached 1000 stem cells in size (which we will denote $p_{1000}(t)$) so that we might calculate the most appropriate time to initiate the follow-up intervention. In figure \ref{figure:turner_mean_variance_survival} \subref{figure:p_exceed} we plot the evolution of $p_{1000}(t)$ over time. It is clear, by $t=100$, that the probability of the tumour having grown to 1000 stem cells, $p_{1000}$ varies significantly depending on the value of $k$ used in the model despite the cells having the same mean CCT\footnote{Note, that by varying $k$ with a fixed cell cycle time we are implicitly varying $\lambda$, the rate of progression through each stage.}. The effects of varying CCTD can clearly be seen to influence the model outcome even in this relatively straightforward linear model of cellular proliferation. In more complex models, in which other species depend in a non-linear manner on the number of cells, the effects will no doubt be further exacerbated. The potential for therapeutic interventions to be based on stochastic mathematical models of cellular proliferation further emphasises the importance of modelling the CCTD correctly.

\subsection{Growth to confluence assays}

Next we investigate the effect of incorporating a more realistic model of cell proliferation on the behaviour of a spatially extended individual-level model of cell migration and proliferation \citep{baker2010cmf}. As such, we alter the mechanism of cellular proliferation from the original, exponentially distributed division times to our more realistic multi-stage Erlang distributed division times and observe the effect this has on the growth of the cell population. In order to achieve this we break the proliferation process into $k$ stages, the passage through each of which has an exponentially distributed waiting time (as described above). As before, we chose the parameter of each stages' waiting time to ensure we have the same mean proliferation attempt time as in the original model.

In more detail, we consider a volume-exclusion process on a regular, square lattice in two dimensions with periodic boundary conditions. Each lattice site, of length $h$, can hold at most one cell. Each repeat realisation begins by initialising, particles uniformly at random across the $L_x\times L_y$ sites of the lattice. Agents can move between adjacent (in the von Neumann sense) lattice sites with rate $P_m$. Movement is unbiased, meaning that once a cell has been chosen to move it does so into one of its four neighbouring lattice sites with equal probability. If the site into which a cell attempts to move is already occupied then that movement event is aborted: the cell attempting movement remains at its current site. 

Agents undergo a proliferation stage change with rate $P_p k$ (giving average rate $P_p$ for unhindered progression through the $k$ stages required for division) this results in the cell's current proliferation stage being incremented by one if the cell is currently in one of the first $k-1$ stages. If the cell is in the final stage (stage $k$) of proliferation and is selected to change stage then the cell attempts to place a daughter in one of its four neighbouring lattice sites with equal probability. If the chosen site is empty, the cell places a daughter in the empty site and the proliferation stages of both the parent and the daughter are reset to unity. However, if the cell attempts to place a daughter in a site which is already occupied then that proliferation event is aborted. In the multi-stage model of the cell cycle, we then have two choices: 
\begin{enumerate}[label=(\arabic*)]
\item the progression stage of the cell attempting proliferation is reset to unity; \label{item:stages_reset}
\item the cell remains in the $k^{th}$ stage.\label{item:stages_held}
\end{enumerate}
In the original model in which $k=1$ these two choices are identical. Under implementation \ref{item:stages_reset} cells would have the same average rate of division attempts as in the original model. However, it could be argued that implementation \ref{item:stages_held} is more realistic as real cells do not reverse through the cell cycle if division is not favourable, but remain held at checkpoints \citep{alberts1994mbc}. We will investigate both possibilities. In order to clearly distinguish the effects of the different CCTDs we will not consider cell death in our simulations. For different values of $P_p$ the population will naturally grow at different rates. As in \citep{baker2010cmf} we will rescale time, $\bar{t}=P_p t$, in order to make population evolutions comparable.

\begin{figure}[h!!!!!!!!!!!!!!!!!!]
\begin{center} 
\subfigure[]{
\includegraphics[width=0.3\textwidth]{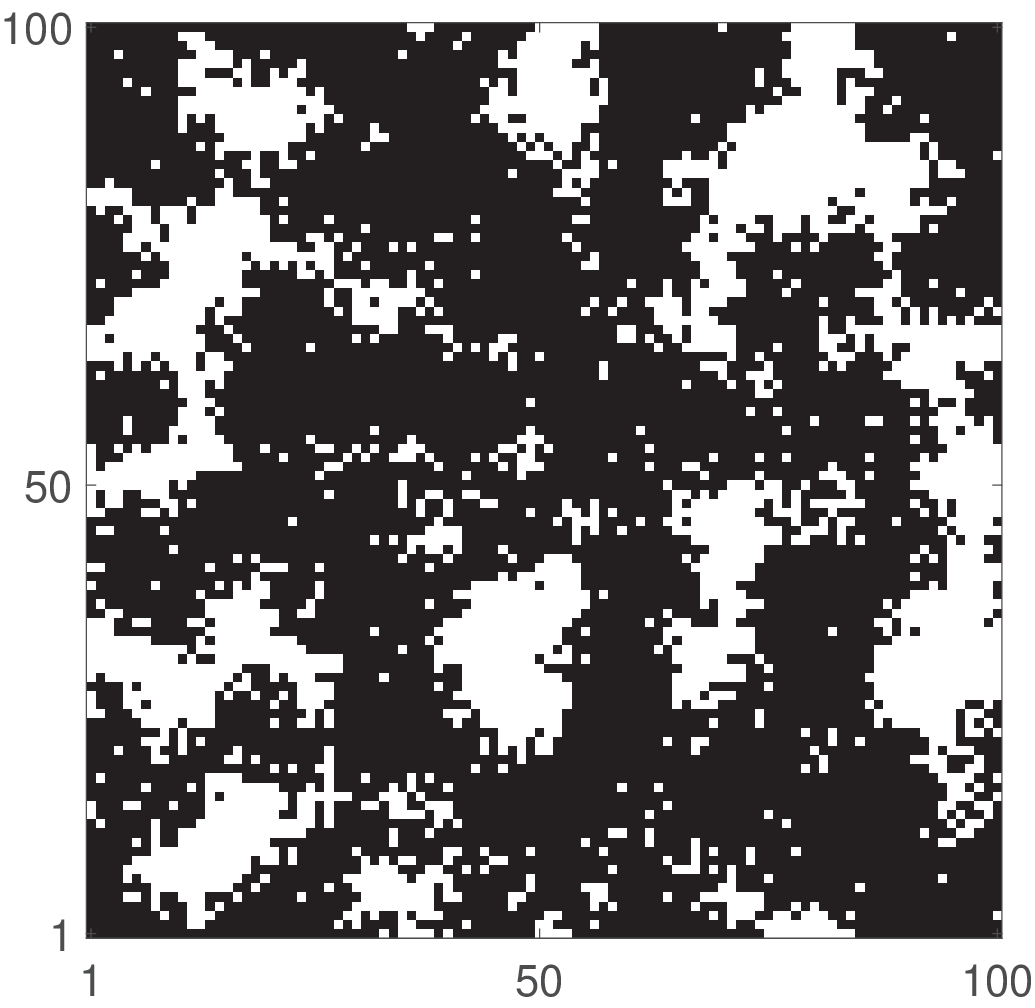}
\label{figure:baker_Pp=1_Pd=0_RESET=1_k=1_t=10}
}
\subfigure[]{
\includegraphics[width=0.3\textwidth]{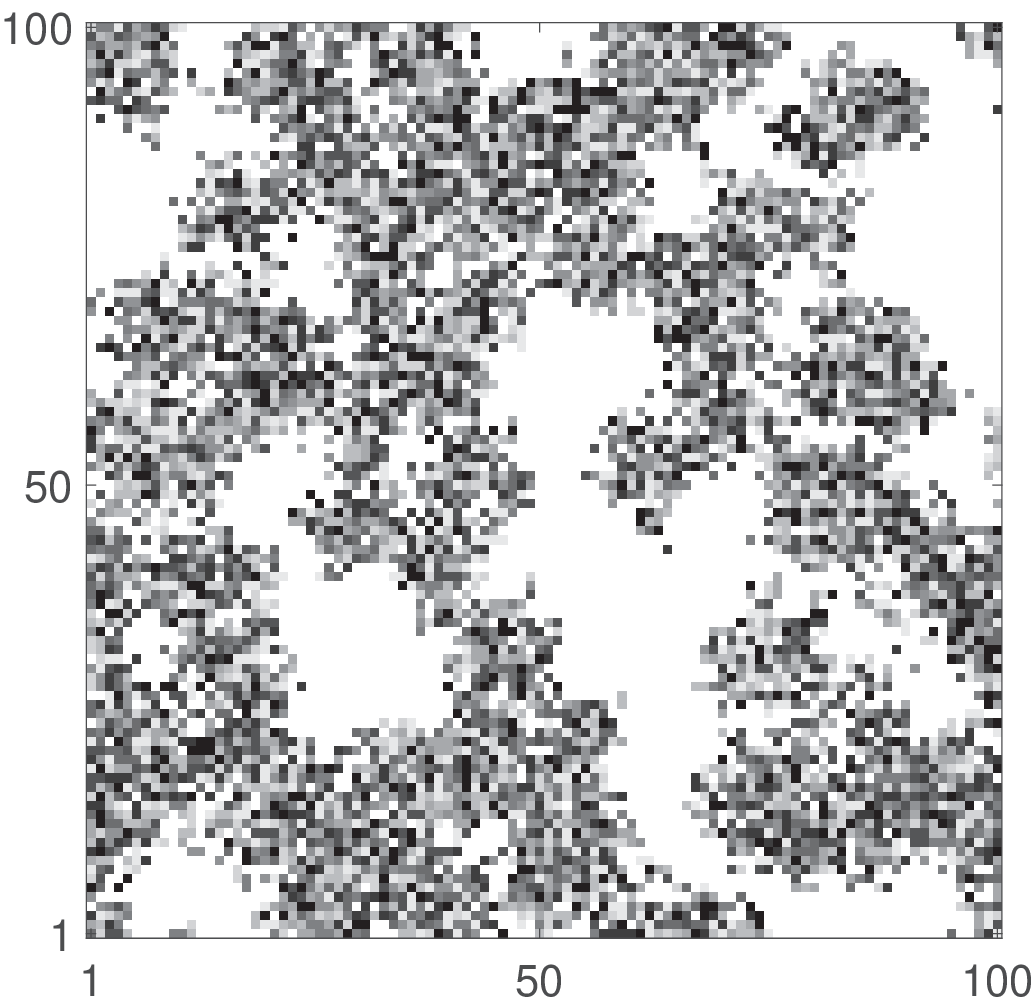}
\label{figure:baker_Pp=1_Pd=0_RESET=1_k=10_t=10}
}
\subfigure[]{
\includegraphics[width=0.3\textwidth]{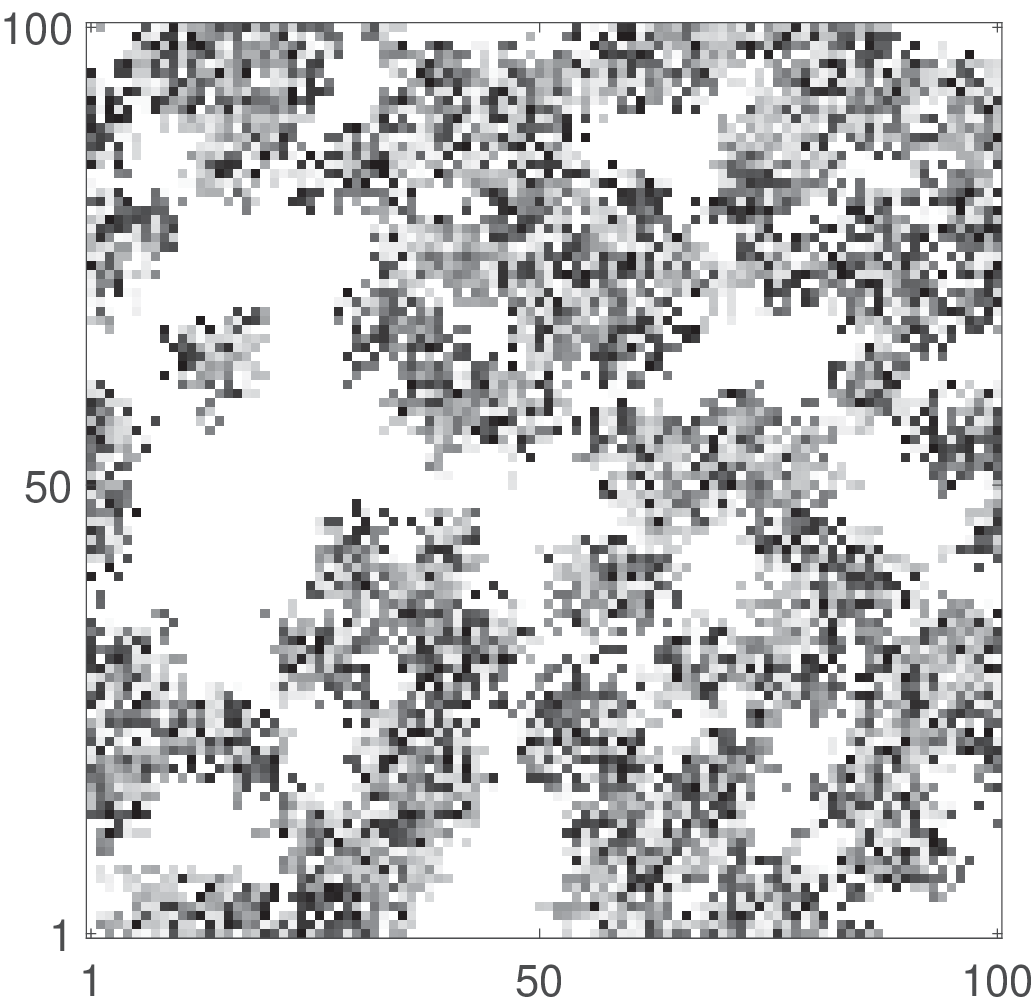}
\label{figure:baker_Pp=1_Pd=0_RESET=1_k=100_t=10}
}

\subfigure[]{
\includegraphics[width=0.3\textwidth]{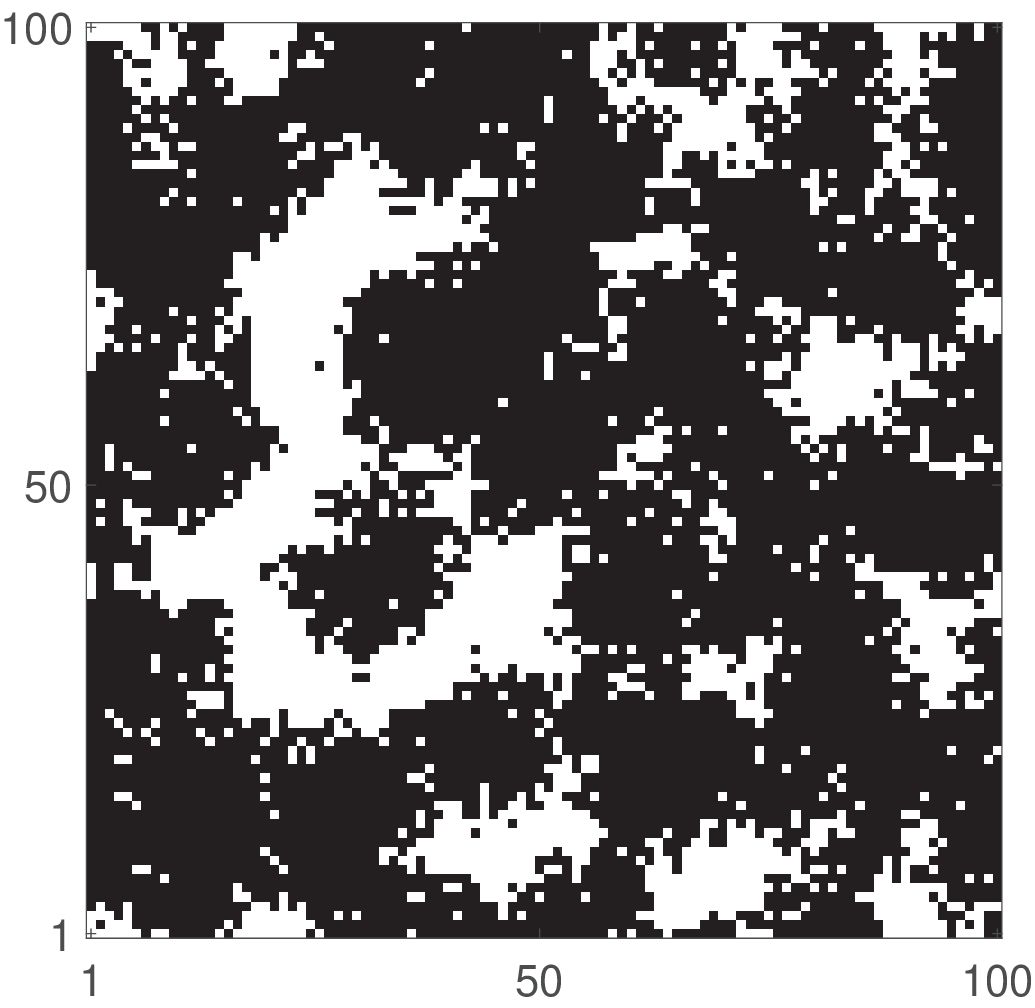}
\label{figure:baker_Pp=1_Pd=0_RESET=0_k=1_t=10}
}
\subfigure[]{
\includegraphics[width=0.3\textwidth]{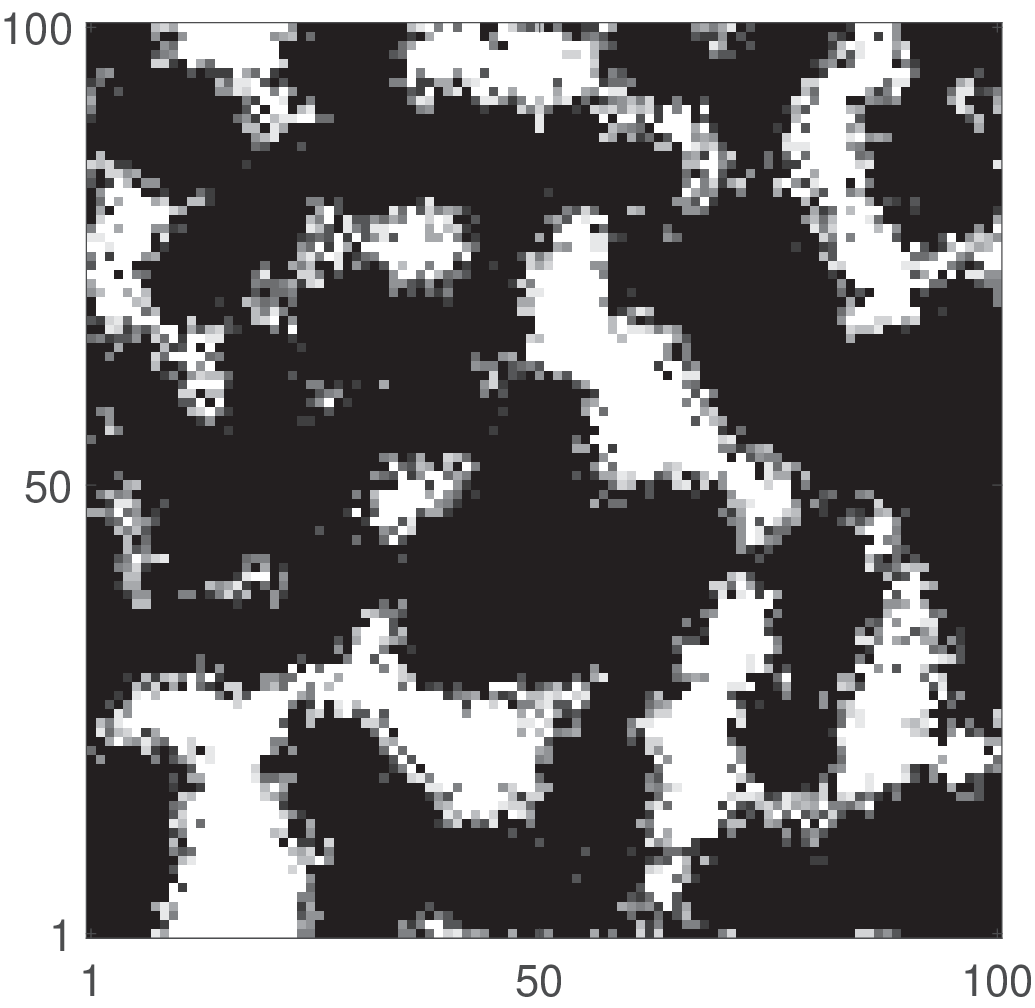}
\label{figure:baker_Pp=1_Pd=0_RESET=0_k=10_t=10}
}
\subfigure[]{
\includegraphics[width=0.3\textwidth]{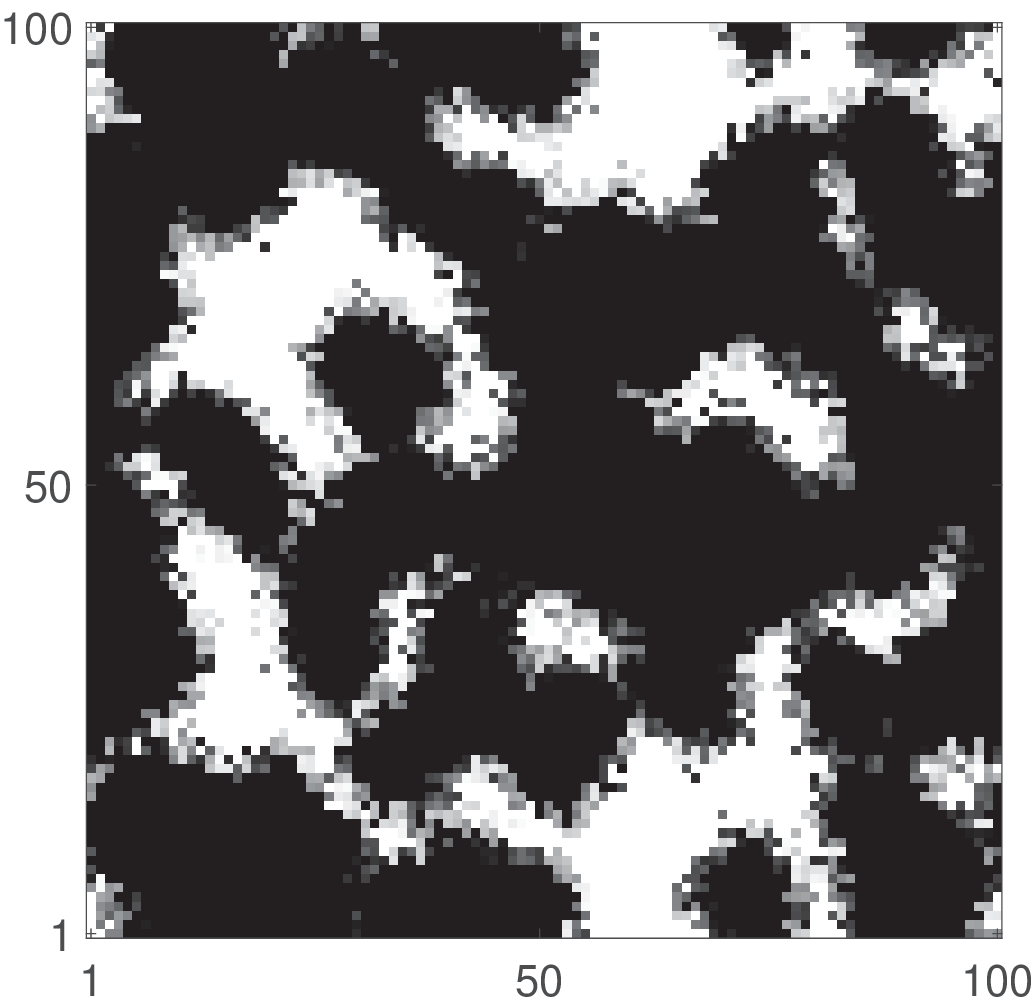}
\label{figure:baker_Pp=1_Pd=0_RESET=0_k=100_t=10}
}
\end{center}
\caption{The influence of the number of proliferation stages, $k$, and the proliferation abortion mechanism on the spatial coverage of cells populating the domain at $t=10$. Cells in different stages are represented by different shades of grey. Darker shading corresponds to later stages and white indicates empty sites. Parameters are $P_m=1$, $P_p=1$ with $L_x=L_y=100$. Initial seeding density is 1\% (\textit{i.e.} 100 cells). Panels \subref{figure:baker_Pp=1_Pd=0_RESET=1_k=1_t=10}--\subref{figure:baker_Pp=1_Pd=0_RESET=1_k=100_t=10} represent implementation \ref{item:stages_reset} for $k=1,10,100$ respectively. Increasing the number of stages in the cell cycle causes a decrease in terminal cell density in this scenario. Panels \subref{figure:baker_Pp=1_Pd=0_RESET=0_k=1_t=10}--\subref{figure:baker_Pp=1_Pd=0_RESET=0_k=100_t=10} represent implementation \ref{item:stages_held} for $k=1,10,100$ respectively. Increasing $k$ causes an increase in the terminal cell density in this scenario.}
\label{figure:baker_density_snapshots}
\end{figure}

Figure \ref{figure:baker_density_snapshots} shows example snapshots of the domain occupancy at rescaled time $\bar{t}=10$ for three different values of $k=1,10,100$. Panels \subref{figure:baker_Pp=1_Pd=0_RESET=1_k=1_t=10}--\subref{figure:baker_Pp=1_Pd=0_RESET=1_k=100_t=10} represent implementation \ref{item:stages_reset} for $k=1,10,100$, respectively. Panels \subref{figure:baker_Pp=1_Pd=0_RESET=0_k=1_t=10}--\subref{figure:baker_Pp=1_Pd=0_RESET=0_k=100_t=10} represent implementation \ref{item:stages_held} for $k=1,10,100$, respectively. Spatial correlations in the occupancies of lattice sites (clusters) are clearly visible in all cases.  

In the multi-stage model, implementation \ref{item:stages_reset} generally leads to less dense colonies than implementation \ref{item:stages_held} since cells do not attempt division as frequently. Under implementation \ref{item:stages_held} (see figure \ref{figure:baker_density_snapshots} \subref{figure:baker_Pp=1_Pd=0_RESET=0_k=10_t=10} and \subref{figure:baker_Pp=1_Pd=0_RESET=0_k=100_t=10}) a clear proliferating rim of (grey) cells can be seen with the bulk of cells being kept at stage $k$ (black). Under implementation \ref{item:stages_reset} every cell can be found in any stage of the cell cycle so it is hard to distinguish the proliferating rim (see figure \ref{figure:baker_density_snapshots} \subref{figure:baker_Pp=1_Pd=0_RESET=1_k=10_t=10} and \subref{figure:baker_Pp=1_Pd=0_RESET=1_k=100_t=10}). The difference between the two implementations however, is not due to aborted proliferation events in the bulk (away from the rim) but to the ability of cells at the proliferating rim to rapidly undergo a further division attempt after an aborted attempt under implementation \ref{item:stages_held}. This suggests that the difference between the two implementations will only be apparent at high densities for which correlations have built up and significant numbers of division attempts are being aborted. 

For low density systems, in which very few particles are adjacent, the mean cell division attempt times are almost the same for all values of $k$ independent of the implementation (\ref{item:stages_reset} or \ref{item:stages_held}). However, the variance in the CCTDs for low density systems affects the rate of growth with larger values of $k$ (less variance in the CCTD) generally leading to slower growth. This effect can be understood by considering equations \eqref{equation:analytical_total_population} and \eqref{equation:t_infinity_limit_total} for a non-excluding population of cells, for which the finite time and asymptotic time behaviours, respectively, of cell populations with different values of $k$ can be contrasted.

\begin{figure}[h!!!!!!!!!!!!!!]
\begin{center} 
\subfigure[]{
\includegraphics[width=0.3\textwidth]{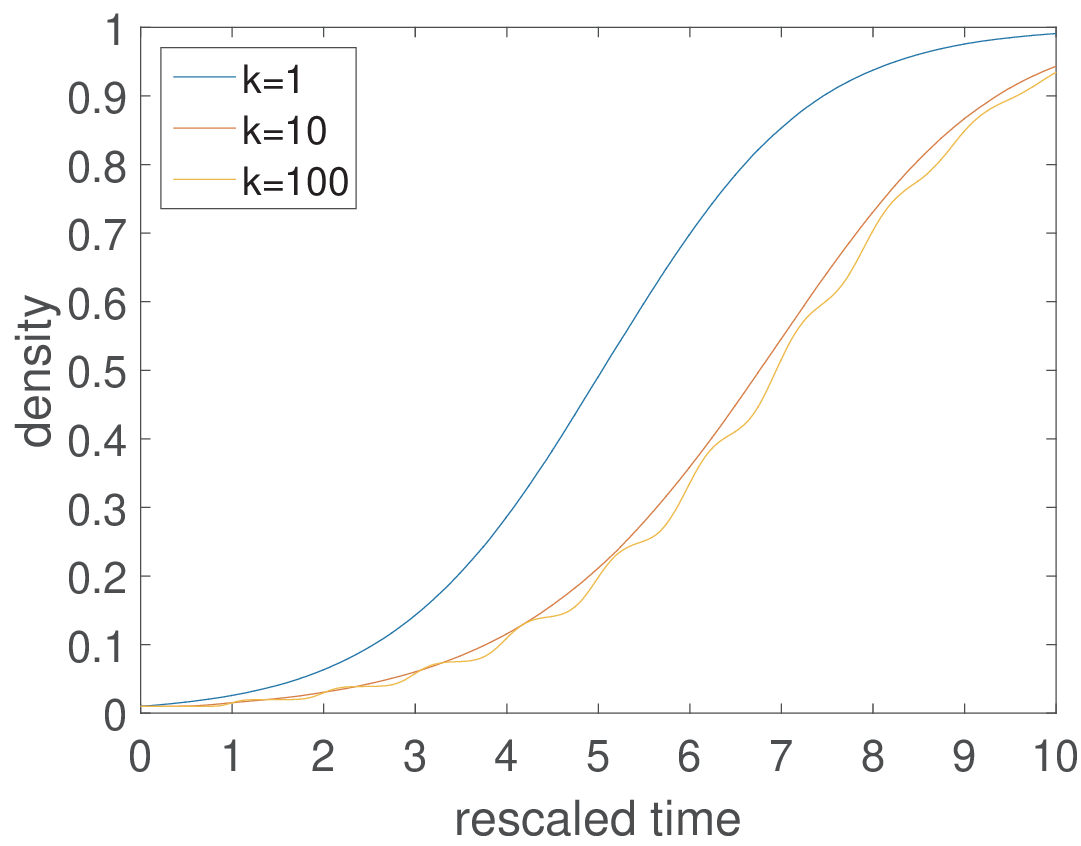}
\label{figure:baker_Pp=0.05_Pd=0_RESET=1_mean_density_evolution}
}
\subfigure[]{
\includegraphics[width=0.3\textwidth]{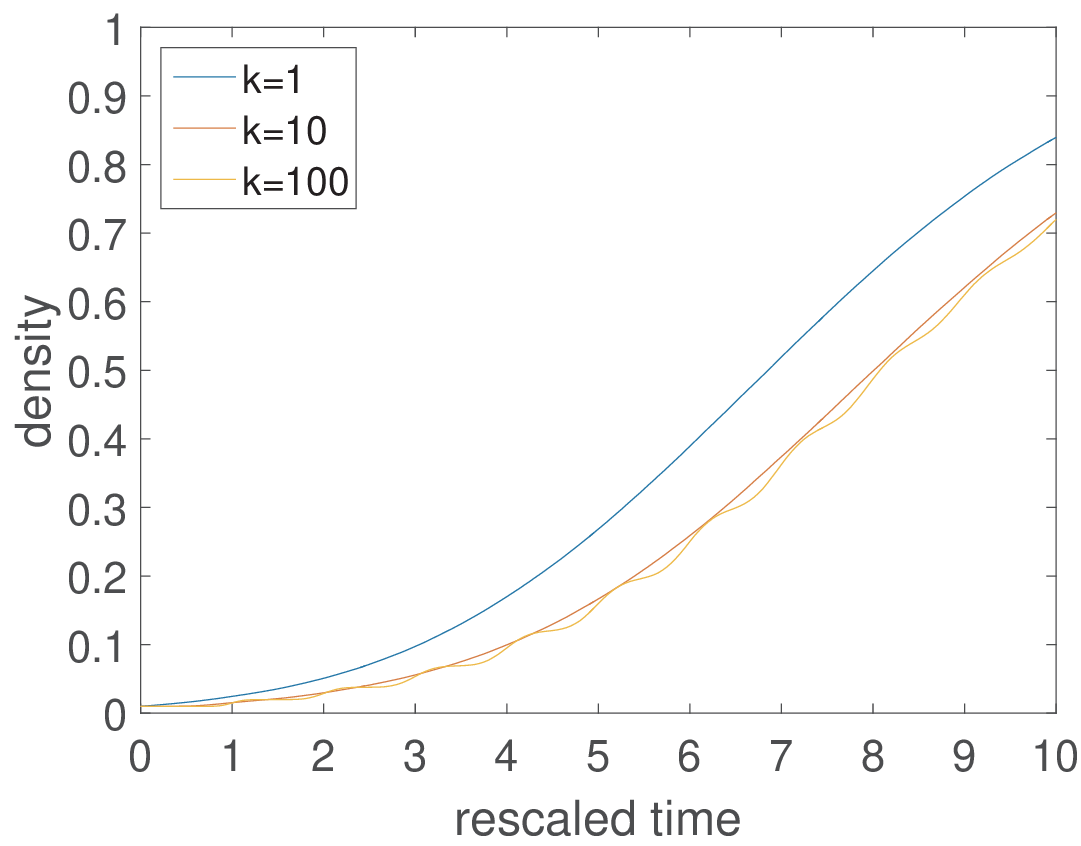}
\label{figure:baker_Pp=0.5_Pd=0_RESET=1_mean_density_evolution}
}
\subfigure[]{
\includegraphics[width=0.3\textwidth]{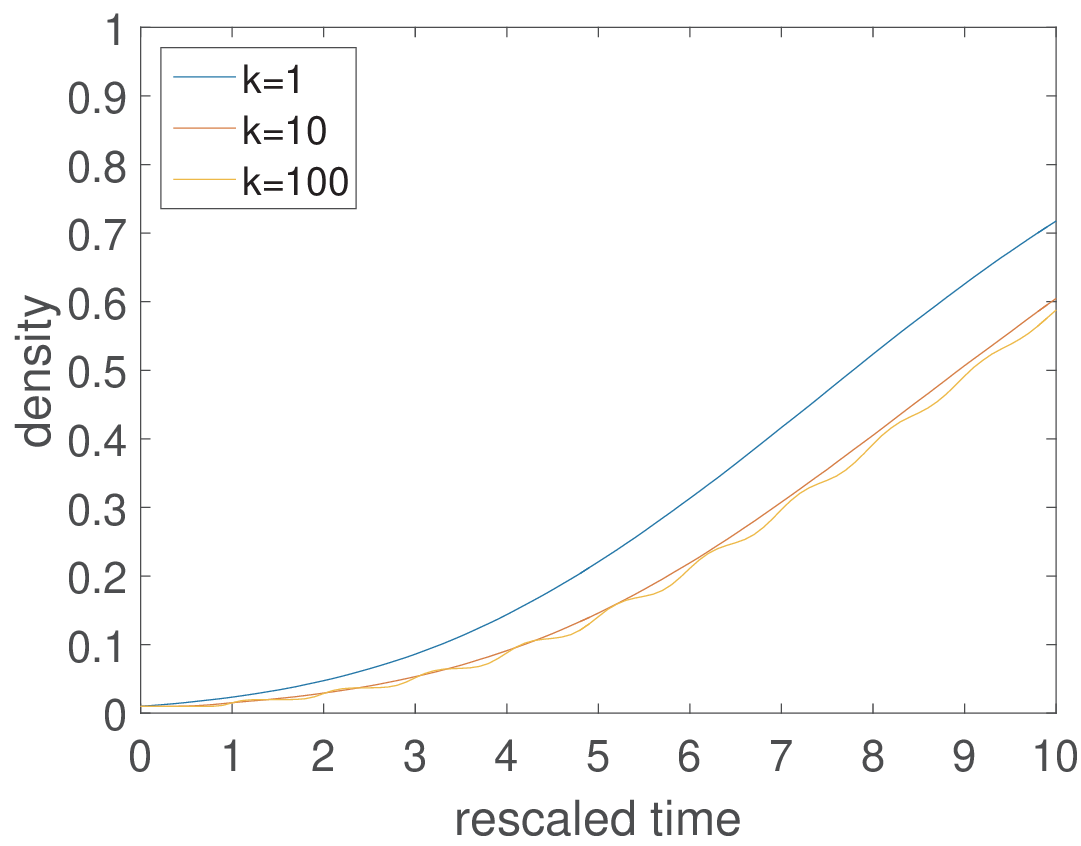}
\label{figure:baker_Pp=1_Pd=0_RESET=1_mean_density_evolution}
}
\subfigure[]{
\includegraphics[width=0.3\textwidth]{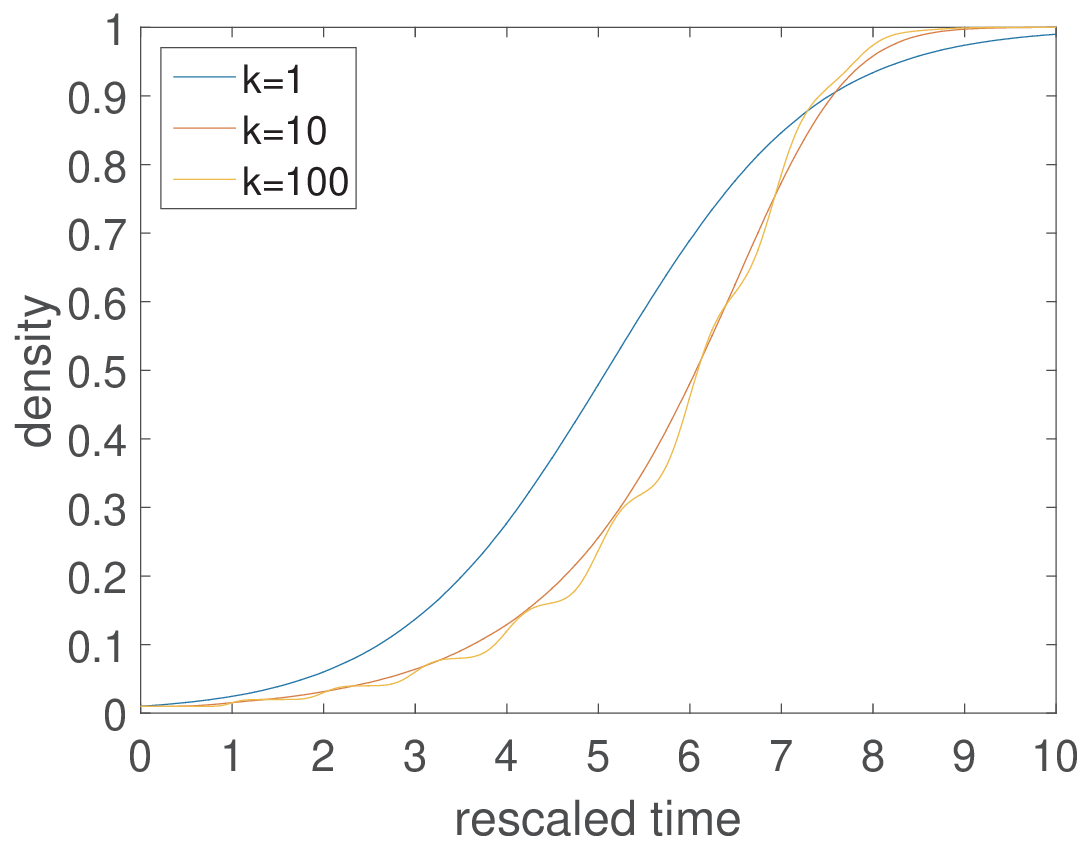}
\label{figure:baker_Pp=0.05_Pd=0_RESET=0_mean_density_evolution}
}
\subfigure[]{
\includegraphics[width=0.3\textwidth]{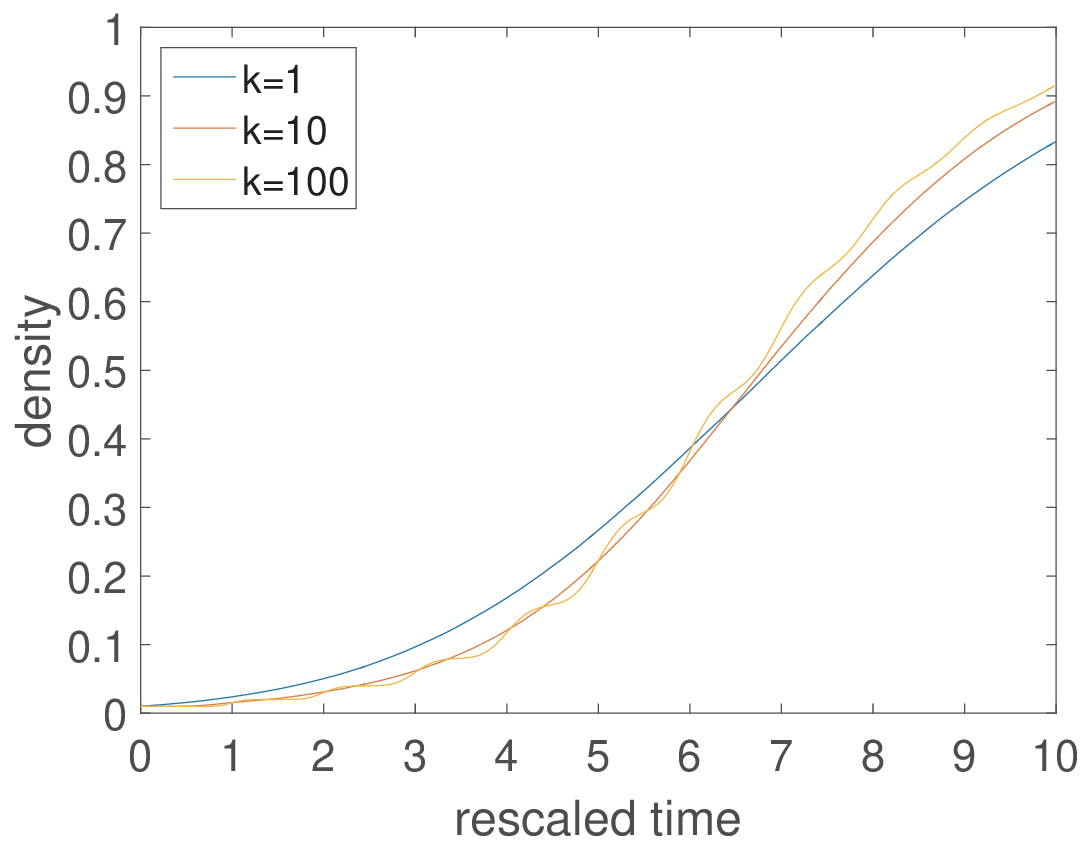}
\label{figure:baker_Pp=0.5_Pd=0_RESET=0_mean_density_evolution}
}
\subfigure[]{
\includegraphics[width=0.3\textwidth]{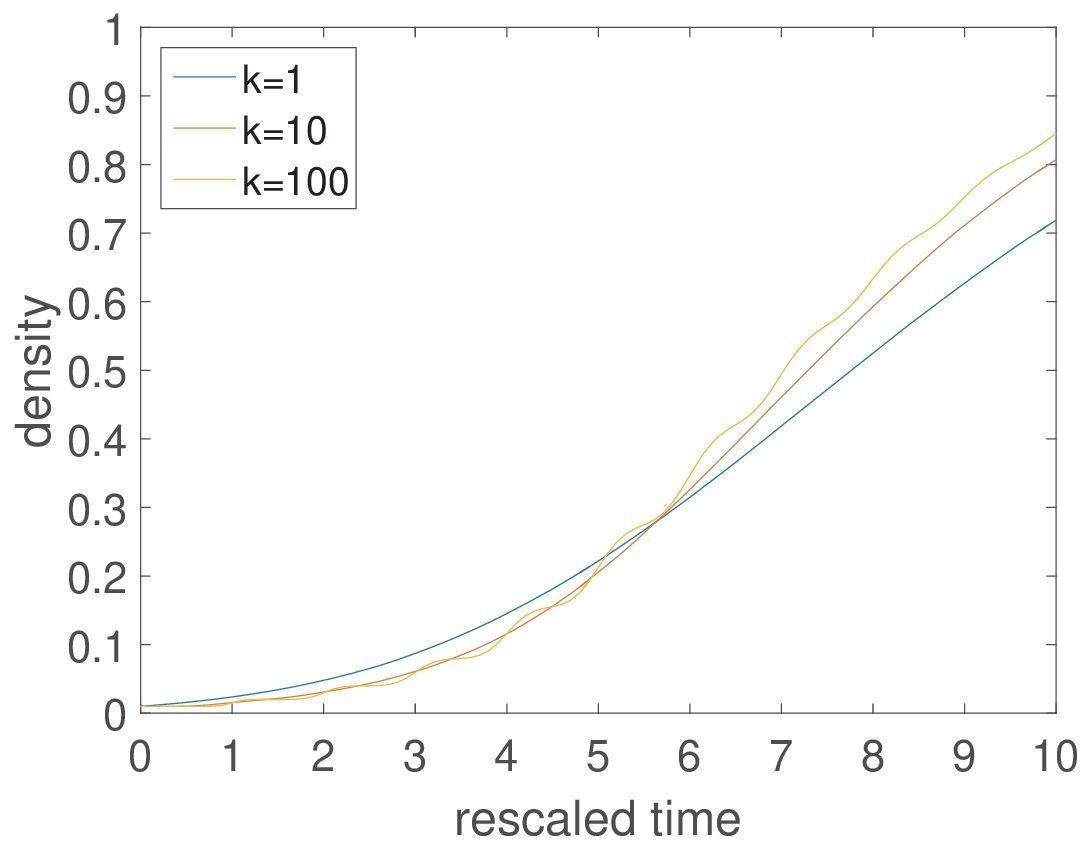}
\label{figure:baker_Pp=1_Pd=0_RESET=0_mean_density_evolution}
}
\end{center}
\caption{The influence of the number of proliferation stages, $k$, and proliferation rate $P_p$ on the evolution of average cell density. Parameters, initial conditions and domain descriptions are as in figure \ref{figure:baker_density_snapshots}. Panels \subref{figure:baker_Pp=0.05_Pd=0_RESET=1_mean_density_evolution}--\subref{figure:baker_Pp=1_Pd=0_RESET=1_mean_density_evolution} represent implementation \ref{item:stages_reset} for $P_p=0.05,0.5,1$ respectively.  Increasing the number of stages in the cell cycle causes growth of cell density to be retarded throughout the simulation. Panels \subref{figure:baker_Pp=0.05_Pd=0_RESET=0_mean_density_evolution}--\subref{figure:baker_Pp=1_Pd=0_RESET=0_mean_density_evolution} represent implementation \ref{item:stages_held} for $P_p=0.05,0.5,1$ respectively. Increasing $k$ causes an initial retardation in growth followed by an acceleration as the effect of density correlations becomes prevalent. Note that figures in which $P_p\neq 1$ are plotted on rescaled time axes for comparison (as described in the main text).}
\label{figure:baker_mean_evolution}
\end{figure}

In figure \ref{figure:baker_mean_evolution} we contrast the evolution of the spatially-averaged density for three values of $k=1, 10, 100$ and three proliferation rates, $P_p=0.05, 0.5, 1$, under implementation \ref{item:stages_reset} (\subref{figure:baker_Pp=0.05_Pd=0_RESET=1_mean_density_evolution}--\subref{figure:baker_Pp=1_Pd=0_RESET=1_mean_density_evolution}) and implementation \ref{item:stages_held} (\subref{figure:baker_Pp=0.05_Pd=0_RESET=0_mean_density_evolution}--\subref{figure:baker_Pp=1_Pd=0_RESET=0_mean_density_evolution}).

Under implementation \ref{item:stages_reset} (see figure \ref{figure:baker_mean_evolution} \subref{figure:baker_Pp=0.05_Pd=0_RESET=1_mean_density_evolution}--\subref{figure:baker_Pp=1_Pd=0_RESET=1_mean_density_evolution}), even as cell-cell correlations build up, multi-stage cells still proliferate more slowly than single-stage cells, since the mean division attempt time remains the same for all values of $k$. The increased variance of cells with fewer stages results in faster population growth.

However, under the more realistic implementation \ref{item:stages_held} (see figure \ref{figure:baker_mean_evolution} \subref{figure:baker_Pp=0.05_Pd=0_RESET=0_mean_density_evolution}--\subref{figure:baker_Pp=1_Pd=0_RESET=0_mean_density_evolution}) cells with multi-stage cell cycles are able to re-attempt division after abortive division events more quickly than they otherwise could under the single-stage cell cycle model. Thus, effective average CCTs for cells with a multi-stage cell cycle at the proliferating rim of a cluster decreases in comparison to cells with a single-stage cell cycle. The faster the  pairwise correlations build up, the more pronounced this effect becomes. With  a very low proliferation rate (in comparison to fixed motility --- see figure \ref{figure:baker_mean_evolution} \subref{figure:baker_Pp=0.05_Pd=0_RESET=0_mean_density_evolution}) cell movement is effective at breaking up correlations meaning that large clusters do not form and that we only see the effect of decreasing mean CCT for larger values of $k$ at late (scaled) times, when the density is higher. Contrastingly, when proliferation is high in comparison to motility (see figure \ref{figure:baker_mean_evolution} \subref{figure:baker_Pp=1_Pd=0_RESET=0_mean_density_evolution}) clusters can form quickly preventing the bulk of cells from proliferating earlier and allowing the cells with multi-stage cell cycles to divide faster on average at the proliferating rim of these clusters than their single-stage counterparts.

It is also worth noting that the greater synchrony in cell division times for larger values of $k$ (exemplified by equations \eqref{equation:k_infinity_limit_total_non_integer}--\eqref{equation:k_infinity_limit_total_integer} for the limit of infinite $k$) is in evidence in the jagged nature of the yellow curves (corresponding to $k=100$) in all six subfigures.

\section{Discussion}\label{section:discussion}
Currently, many stochastic models which incorporate cell proliferation employ the ubiquitous Gillespie stochastic simulation algorithm \citep{gillespie1976gmn,gillespie1977ess}. Unfortunately, in its basic form the Gillespie algorithm represents all events as exponentially distributed. Cell cycle times are not exponentially distributed and can not, therefore be represented by a single reaction event in the Gillespie algorithm. Modelling cell cycle times as a single exponentially distributed event can lead to significant alterations in model behaviour in comparison to more appropriate CCTDs. Consequently, we postulated a simple, general hypoexponentially distributed CCT which can be broken down into exponentially distributed stages allowing for straightforward simulation with the popular Gillespie algorithm. To account for ease of parameter identification we suggested two special cases of this more general model which have been shown to do an excellent job of recapitulating CCTDs \citet{golubev2016aie}.

We postulate that the general hypoexponential distribution \citep{zhou2007kas} or even the more specific Erlang \citep{gibson2000ees,svoboda1994fam} or exponentially modified Erlang \citep{lucius2003gma} inter-event distribution time models could be used to allow the simplified simulation of complex biochemical and biophysical processes (\textit{e.g.} enzymatic reactions \citep{nelsen2005pb}, allosteric transitions in ion channels \citep{qin2004mbf}, the movement of molecular motors \citep{schnitzer1995skp}, DNA unwinding \citep{lucius2003gma}) using the Gillespie algorithm. More generally, non-Markovian processes for which only the inter-event distribution, rather than the mechanism which generates this distribution, is important might be simulated efficiently using our proposed mechanism \citep{gibson2000ees,floyd2010aki,lucius2003gma,zhou2007kas}.

We employed our improved model of cell cycle proliferation times on two recent models of real biological processes \citep{turner2009cbc,baker2010cmf}. In each case we found that the incorporation of multiple stages to the cell cycle led to significant differences in the population size in comparison to the original exponentially distributed CCT model. We suggest that these difference will hold more generally throughout stochastic models in which CCTs are currently modelled as exponential. In particular, we intend to investigate the effects of our modified CCTD on the speed of invasion of a population of migrating and proliferating cells.

The application here of hypoexponentially distributed CCTs built up from a number of intermediary exponential stages assumes that the CCT is not correlated between direct descendants or within a given generation. Whilst there are scenarios in which there is no evidence for a correlation in CCT between related cells \citep{schultze1979ttt} there are other situations in which this assumption is clearly invalid \citep{duffy2012aib,hawkins2009scp}. It is possible that some of these correlation effects can be attributed to the environment in which the cells are proliferating. However, in NIH 3T3 cells a clear correlation has been observed between daughter cells of a given mitotic event compared to more distant relatives; implying a heritable predisposition \citep{mort2014fbc}. Therefore, one obvious extension to this work would be to incorporate the effects of correlations in cell and phase times to better reflect the biological heterogeneity of a given system.


\section{Appendix A - Materials and Methods}

In order to determine cell cycle times in cell culture, NIH 3T3 Flp-In cells \citep{mort2014fbc} were seeded at various densities in phenol-red free dulbeccos modified eagle medium (DMEM) containing 10\% fetal calf serum, 1\% Penicillin/Streptomycin and 100 \textmu g/ml  Hygromycin B on glass bottomed 24-well culture plates (Greiner bio-one, UK). The next day, time-lapse imaging was performed on subconfluent wells with a 20$\times$ objective using a Nikon A1R inverted confocal microscope in a heated chamber supplied with 5\% CO$_2$ in air. The time elapsed between mitotic events was measured using the Fiji \citep{schindelin2012fos} distribution of ImageJ -  an open source image analysis package based on NIH Image \citep{schneider2012nih}.

\section{Appendix B - Psuedocode for the multi-stage model of the cell cycle}\label{section:appendix_B_pseudocode}
One of the original (and most popular) implementations of the mathematically exact SSA is known as the direct method \citep{gillespie1977ess}. Here we present pseudocode for the direct method implementation of the simple multi-stage model of the cell cycle (corresponding to Erlang distributed CCTs) in a well mixed context. 

Let $\bs{X}(t)$ be a vector of length $k$ which contains the number of cells in each stage at time $t$. A time interval $\tau$, until the next stage advancement event, is generated. Along with it, an index $j$, is chosen which determines from which stage a cell will advance from time $t+\tau$. The changes in the numbers of cells caused by the stage advancement are implemented, the propensity functions (progression rate, $\lambda$, multiplied by the number of cells in each stage) are altered accordingly and the time is updated, ready for the next ($\tau,j$) pair to be selected.
A method for the implementation of this algorithm is given below:
\begin{enumerate}
\label{algorithm:original_DM}
 \item Initialize the time $t=t_0$ and the number of cells in each stage, $\bs{X}(t_0)=\bs{x}_0$.
 \item \label{SSAalg:evaluatepropfunctions} Evaluate the propensity functions, $a_j(\bs{X}(t))=\lambda X_j(t)$, (for $j=1,\dots,k$) associated with the advancement of cells from their respective stages, and their sum $a_0(\bs{X}(t))=\sum^k_{j=1}a_j(\bs{X}(t))$.
\item  Generate two random numbers $rand_1$ and $rand_2$ uniformly distributed in
$(0,1)$.\label{step:generate_two_random_numbers}
\item Use $rand_1$ to generate a time increment, $\tau$, an exponentially distributed random variable with mean $1/a_0(\bs{X}(t))$. \textit{i.e.}
\begin{equation}
\tau=\frac{1}{a_0}\ln\left(\frac{1}{rand_1}\right)\label{equation:time-stepchoice}.\nonumber
\end{equation}
\label{step:time-step_generation}
\item Use $rand_2$ to generate index of the next stage advancement event, $j$, with probability $a_j(\bs{X}(t))/a_0(\bs{X}(t))$, in proportion with its propensity function. \textit{i.e.} find $j$ such that\footnote{Note in step \ref{step:reaction_choice}, that in the case $j=1$ we assume $\protect\sum_{j'=1}^{0}=0$\label{footnote:1}.}
\begin{equation}
 \displaystyle\sum_{j'=1}^{j-1}a_{j'}(\bs{X}(t))<a_0(\bs{X}(t))\cdot rand_2<\displaystyle\sum_{j'=1}^{j}a_{j'}(\bs{X}(t)).\nonumber
\end{equation}
\label{step:reaction_choice}
\item Update the time, $t=t+\tau$, and the state vector to reflect the advancement of one cell from the chosen stage, 
\begin{align}
 X_j=X_j-1 \quad &\text{if}\quad j = 1,\dots,k\\
 &\text{and}\\
X_{j+1}=X_{j+1}+1 \quad &\text{if}\quad  j\neq k, \\
&\text{or}\\
X_1=X_1+2\quad  &\text{if}\quad j = k. 
\nonumber \label{equation:update_cell_numbers}
\end{align}

\item If $t<t_{final}$, the desired stopping time, then go to step (\ref{SSAalg:evaluatepropfunctions}). Otherwise stop.
\end{enumerate}

\section*{Acknowledgements}
This collaboration was supported by an LMS research in pairs grant (Grant \#41461).
Dr Richard Mort is supported by funding from the NC3Rs and Medical Reserach Scotland (Grant \#NC/K001612/1 and \#NC/M001091/1).
Dr Christian Yates would like to thank the CMB/CNCB preprint club for constructive and helpful comments on a preprint of this paper. 


\end{document}